\documentclass[letterpaper,leqno]{IEEEtran}
\pdfoutput=0
\usepackage{mystyle}

\newcommand{\Avg}{\mathop{\op{Avg}}}

\title{Success Exponent of Wiretapper:\\ A Tradeoff between Secrecy and Reliability}

\author{Chung Chan
  \thanks{Manuscript written on May~28, 2008.}
  \thanks{Chung Chan
    (\href{mailto:chungc@mit.edu}{chungc@mit.edu}) is with the
    Laboratory of Information and Decision Systems, Department of
    Electrical Engineering and Computer Science, Massachusetts
    Institute of Technology.}}
\date{May~23, 2008}
\begin{document}

\maketitle
\begin{abstract}
  Equivocation rate has been widely used as an information-theoretic
  measure of security after Shannon\cite{shannon1948}. It simplifies
  problems by removing the effect of atypical behavior from the
  system. In \cite{merhav1999}, however, Merhav and Arikan considered
  the alternative of using guessing exponent to analyze the Shannon's
  cipher system. Because guessing exponent captures the atypical
  behavior, the strongest expressible notion of secrecy requires the
  more stringent condition that the size of the key, instead of its
  entropy rate, to be equal to the size of the message.\footnote{This
    is the condition for a finite system to achieve perfect secrecy as
    pointed out by Shannon\cite{shannon1948}.} The relationship
  between equivocation and guessing exponent are
  also investigated in \cite{malone2005}\cite{massey1994} but it is
  unclear which is a better measure, and whether there is a
  unifying measure of security.
  
  Instead of using equivocation rate or guessing exponent, we study
  the wiretap channel in \cite{csiszar1978} using the success
  exponent, defined as the exponent of a wiretapper successfully learn
  the secret after making an exponential number of guesses to a
  sequential verifier that gives yes/no answer to each guess. By
  extending the coding scheme in \cite{csiszar1978}\cite{korner1980}
  and the converse proof in \cite{dueck1979} with the new Overlap
  Lemma~\ref{lem:overlap}, we obtain a tradeoff between secrecy and
  reliability expressed in terms of lower bounds on the error and
  success exponents of authorized and respectively unauthorized
  decoding of the transmitted messages. From this, we obtain an inner
  bound to the region of strongly achievable public, private and
  guessing rate triples for which the exponents are strictly positive.
  The closure of this region is equivalent to the closure of the
  region in Theorem~1 of \cite{csiszar1978} when we treat equivocation
  rate as the guessing rate. However, it is unclear if the inner bound
  is tight.
\end{abstract}


\section*{Acknowledgment}
\label{sec:acknowledgment}

I would like to thank Professor Lizhong Zheng for his guidance and
valuable comments; Jing Sheng for his inspiring discussion that
eventually leads to the proof of Proposition~\ref{pro:rates}; and above
all, my family for their support and encouragements.

\section{Introduction}
\label{sec:introduction}


The basic model of a cryptographic/secrecy system involves a sender
Alice who wants to send a message $\RS$ \emph{as secretly as
  possible} to the intended receiver Bob. The basic model of a
cryptanalytic attack, on the other hand, involves a
cryptanalyst/wiretapper Eve who attempts to learn the secret \emph{as
  much as possible} based on her observation $\RZ$. How secretly
a message is sent, or how much information is leaked, must therefore
be quantified before one can design and optimize a cryptographic
system or a cryptanalytic attack for the respective purposes.

The aposteriori probability function $P_{\RS|\RZ}$ is a
sufficient statistics of the security of the system as it gives all
the possible values of the secret and their associated probabilities
for every possible realization of the wiretapper's observation. In
particular, the important notion of a system being perfectly secure,
referred to as \emph{perfect secrecy} by Shannon\cite{shannon1948},
can be characterized as the aposteriori probability equal to the
prior, i.e.\ $P_{\RS|\RZ}=P_{\RS}$. In other words, Eve's observation
is independent of the secret, or equivalently, the system is at the
same level of security whether $\RZ$ is observed or not.

\begin{figure}
  \centering
  \input{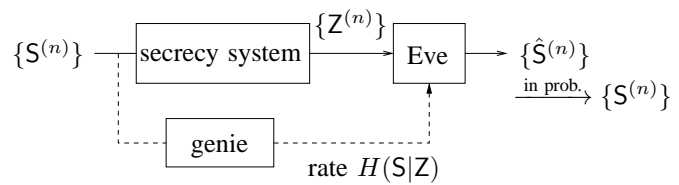}
  \caption{Genie-aided correction channel}
  \label{fig:equiv}
\end{figure}

It is convenient to summarize the aposteriori probability function by
the index called \emph{equivocation} $H(\RS|\RZ)$. It is roughly the
amount of information the wiretapper needs to gather in addition to
$\RZ$ to perfectly recover $\RS$. One precise operational meaning of
equivocation, as illustrated in \figref{fig:equiv}, is the minimum
achievable rate for source coding an iid sequence of $\RS^{(n)}$ with
the iid sequence of $\RZ^{(n)}$ as side information at the
decoder.\footnote{This is the correction data model originally
  proposed by Shannon\cite{shannon1948} except that the genie does not
  need to know $\RZ$ nor any decision feedback from Bob.} To achieve
perfect secrecy, it is necessary and sufficient to have
$H(\RS|\RZ)=H(\RS)$. Alice can also try to
protect the secret up to an equivocation $H(\RS|\RZ)$ below $H(\RS)$
if perfect secrecy is costly and unnecessary.

The amount of additional information Eve needs to gather to break the
system may not reflect how difficult it is to obtain them. For
example, getting just one bit of information from Alice or someone who
know the secret may require significant effort in the search for that
person, followed by lengthy interrogation. In some situations, Eve
does not play a passive role of receiving additional information that
is concisely stated (i.e.\ maximally compressed by a genie), but
instead plays an active role in identifying and extracting relevant
information from disorganized sources. Thus, one should question
whether equivocation is applicable for the case of interest, albeit
its mathematical convenience.

A natural alternative measure of security, as investigated by Merhav
and Arikan\cite{merhav1999}, is roughly the \emph{ability} that Eve
perfectly learn the secret from yes/no answers to ``Is the secret
equal to ...?'' type of questions. In the model, Eve sequentially
verify her guesses of the secret by asking yes/no questions. The
number of guesses and verifications she needs to make until she is
within some probability of guessing the secret correctly indicates her
effort and ability to extract information about the secret. Sometimes
the system itself provides such a verifier which help correct careless
mistakes made by the authorized user. This potentially leaks
information to unauthorized users who also have access to the
verifier, just as in the case of a login system. As a system designer,
he may be interested to know how many wrong passwords should be
allowed for each session so that the chances of successfully breaking
into the account is reasonably small. Although this success
probability does not have a way to express the notion of perfect
secrecy in general (See Example~\ref{eg:perfect}), it is a natural fit
for this problem as it provides the number of trials as an additional
parameter to optimize.

In the sequel, we will consider the wiretap channel problem in
\cite{csiszar1978}. A key result from \cite{csiszar1978} is the single
letter characterization of the secrecy capacity, defined as the
maximum rate at which the secret can be transmitted to Bob by a block
coding scheme with arbitrarily small error probability and the
equivocation rate equal to the message rate. Transmitting at rate
above this secrecy capacity, one faces the trade-off a lower
equivocation rate. Transmitting at rate below the secrecy capacity,
however, equivocation rate is capped at the message rate. There seems
to be little point in further reducing the rate below secrecy
capacity. If one also cares about delay, i.e. how fast the error
probability converges to zero, further reducing the rate below secrecy
capacity can be beneficial. What is the tradeoff then?

Secrecy comes with a cost of \emph{reliability} of the authorized
decoding. To characterize which level of secrecy and reliability are
simultaneously achievable for each rate, we will use the
standard notion of \emph{error exponents} for Bob and Eve in decoding
their messages as a measure of reliability.  For secrecy, we will use
the exponent of the success probability, or \emph{success exponent}
for short, that Eve learns the secret within an exponential
number of guesses.

The rest of the paper will be organized as follows.
Section~\ref{sec:problem-formulation} defines the wiretap channel
problem we consider.
Section~\ref{sec:coding-scheme} describes the
proposed coding scheme.
Section~\ref{sec:success-exponent} explains the computation of the
success exponent using a technique we call the Overlap
Lemma~\ref{lem:overlap}.
Section~\ref{sec:error-exponents} explains the computation of the
error exponents using the Packing
Lemma\cite{csiszar1981}.
Finally, the desired lower bounds on the exponents will be stated in
Section~\ref{sec:result}. 
Section~\ref{sec:conclusion} gives the conclusion and some open
problems. For readers who would like to skip to the main result,
Section~\ref{sec:preliminaries} provides a brief summary of notations.

\section{Preliminaries}
\label{sec:preliminaries}
Calligraphic font denotes a set, e.g.\ $\setA$, which is always
assumed finite unless otherwise stated.  $2^{\setA}$ and $\setA^c$
denote the power set and complement of $\setA$ respectively.
$\setA\cup\setB$, $\setA\cap\setB$ and $\setA`/\setB$ denotes the
usual set operations, which are the union, intersection, and
difference respectively. $\Avg_{a\in\setA}$ (or $\Avg_a$ for short)
denote the averaging operation $\frac1{\abs\setA}\sum_{a\in\setA}$.
$`R$, $`R_+$ and $`Z^+$ denotes the set of real numbers, non-negative
real numbers, and positive integers. Occasionally without
ambiguity, a positive integer $L$ will also be used to denote the set
$\Set{1,\dots,L}$ as in $l\in L$. Bold letter such as $\Mx$ denotes an
$n$-sequence $\Set{x^{(i)}}_{i=1}^n=(x^{(1)},\dots,x^{(n)})$; and
$\Mu\circ\Mx$ denotes element-wise concatenation
$\Set{(u^{(i)},x^{(i)})}_{i=1}^n$.

San serif font is used for random variables and stochastic functions,
e.g.\ $\RX$, $\Rf$ and $\RW_b$. $\rsfsP(\setY)^\setX$ denotes the set
of all possible conditional probability distributions $P_{\RY|\RX}$ of
a random variable $\RY$ taking values from $\setY$, denoted as
$\RY\in\setY$, given a random variable $\RX\in\setX$. The
(conditional) probability distribution will also be viewed as a row
vector (matrix). e.g.\ $P_\RX P_{\RY|\RX}$ denotes the matrix
multiplication, which gives the marginal distribution $P_\RY$.
$P_\RX\circ P_{\RY|\RX}$ denotes the direct product, which gives the
joint distribution $P_{\RX,\RY}$ of the pair $(\RX,\RY)$ in this
case. $P_\RX^n$ denotes the $n$-th direct product such that
$P_{\RX}(\Mx)=\prod_{i=1}^n P_\RX(x_i)$. For any subset $\setA\subset
\setX$, $P_{\RX}(\setA)=\sum_{x\in\setA} P_{\RX}(x)$. $\opE(\RX)$
denote the expectation of $\RX$. $`d_{\op{var}}(P,Q)$ denotes the
variation distance \eqref{eq:var} between $P$ and $Q$.

Following the notations in \cite{csiszar1981} for the \emph{method of
  types}, $P_{\Mx}$ and $P_{\My|\Mx}$ denotes the type~\eqref{eq:type}
and respectively canonical conditional type~\eqref{eq:canonical}.
`Canonical' refers to the constraint (for convenience) that
$P_{\My|\Mx}(y|x)=1/\abs\setY$ if $P_{\Mx}(x)=0$ for all
$(x,y)\in\setX\times\setY$. $T_Q^{(n)}$ or $T_Q$ for short denotes the
class of $n$-sequences of type $Q$.  $T_{V}(\Mx)$ denotes the
$V$-shell of $\Mx$. $\rsfsP_n(\setX)$ denotes the set of all types for
sequences in $\setX^n$.  $\rsfsV_n(Q,\setY)$ ($\rsfsV_n(Q)$ or
$\rsfsV_n$ for short) denotes the set of all canonical conditional
types $V$ for sequences in $\setY^n$. $I(Q,V|P)$, $D(V\|W|Q)$, and
$H(V|P)$ are the conditional mutual information~\eqref{eq:cI}, divergence
\eqref{eq:D} and entropy~\eqref{eq:H} respectively.  $I(\Mx \wedge
\My)$~\eqref{eq:eI} denotes the empirical mutual
information. Equivalently, we write $T_{\RX}:=T_{P_{\RX}}$ and
$T_{\RY|\RX}:=T_{P_{\RY|\RX}}$, which are non-empty if the corresponding
distributions are valid (conditional) types. $\abs{T_{\RY|\RX}}$
denote $\abs{T_{P_{\RY|\RX}}(\Mx)}$ with $\Mx\in T_{\RX}$.

To express inequality in the exponent for functions in $n$, we use
$a_n \dotleq b_n$ to denote $\limsup_{n\to`8} \frac1n \log a_n$ is no
larger than $\liminf_{n\to`8} \frac1n \log b_n$.
A piecewise function 
will be expressed in terms of $\abs{a}^+:=\max\Set{0,a}$ and
$\abs{a}^-:=\min\Set{0,a}$.

\section{Problem formulation}
\label{sec:problem-formulation}

\subsection{Transmission model}
\label{sec:transmission-model}


\begin{figure}
  \centering
  \input{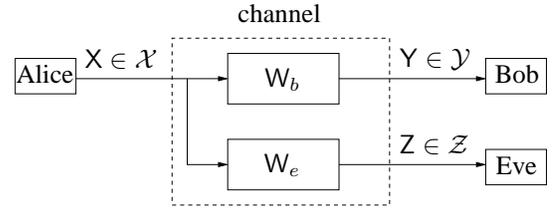}
  \caption{Wiretap channel model}
  \label{fig:wiretap}
\end{figure}

\figref{fig:wiretap} illustrates a single use of the discrete
memoryless wiretap channel $(W_b,W_e)$ using the dummy random variables $\RX$,
$\RY$ and $\RZ$. Alice sends a random variable $\RX$ through the channel.
$P_{\RX}\in \rsfsP(\setX)$ is the probability distribution
function/vector of $\RX$ over the finite set $\setX$, such that
$P_{\RX}=\Pr\Set{\RX=x}\;(x\in\setX)$ and
$P_{\RX}(\setA)=\Pr\Set{\RX\in\setA}\;(\setA\subset\setX)$.

The channel is denoted by the pair $(W_b
\in\rsfsP(\setY)^\setX,W_e\in\rsfsP(\setZ)^\setX)$ of
conditional probability distributions. We write $\RW_b(\RX)$ and
$\RW_e(\RX)$ as the channel output $\RY$ and resp. $\RZ$ observed by Bob
and resp.\ Eve. The conditional distribution
$P_{\RY|\RX}(y|x):=\Pr\Set{\RY=y|\RX=x}$ equals $W_b(y|x)$ for all
$(x,y)\in\setX\times\setY$, and similarly for $P_{\RZ|\RX}$. 
For the case of interest, all sets $\setX$, $\setY$ and
$\setZ$ are finite and the correlation between $\RY$ and $\RZ$ given
$\RX$ need \emph{not} be specified.

\begin{figure}
  \centering
  \input{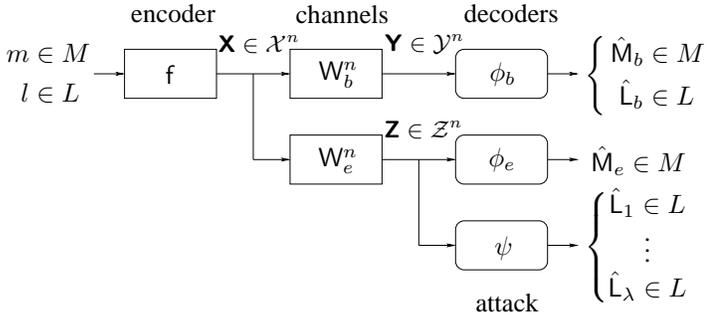}
  \caption{Transmission model}
  \label{fig:transmission}
\end{figure}

To transmit information through this channel, we will consider the
(data) transmission model illustrated in \figref{fig:transmission}
with \emph{block length} $n$. Following \cite{csiszar1978}, we
consider $n$ uses of the channel with \emph{stochastic encoding}, and
deterministic decoders at the receivers. As pointed out in
\cite{csiszar1978}, stochastic encoding, i.e.\ randomization in the
encoder \emph{during transmission}, increases secrecy by adding noise
as a \emph{physical barrier} to eavesdropping while deterministic
decoding does not lose optimality for the case of interest.

As shown in \figref{fig:transmission}, Alice chooses a
\emph{public/common message} $m$ out of a set of $M$ possible messages
to convey to both Bob and Eve, and a \emph{private/secret/confidential
  message} $l \in L$ only to Eve. ($l\in L$ is a short-hand notation
for $l\in\Set{1,\dots,L}$.) Since the message $m$ for Eve is a
degraded version of the message $(m,l)$ to Bob, this is identical to
the \emph{asymmetric broadcasting} of \emph{degraded message
  sets}\cite{csiszar1981} except for the additional secrecy concern.

In the \emph{transmission phase}, Alice first passes the message
through a stochastic encoder denoted by the conditional probability
distribution $f \in \rsfsP(\setX^n)^{M \times L}$. We write
$\Rf(m,l)$ as the output codeword, which is denoted by the dummy random
$n$-sequence $\RMX:=\Set{\RX^{(i)}}_{i=1}^n$ in
\figref{fig:transmission}. The encoder can be viewed as an
artificial channel, through which the output codeword $\RMX$ of the
message $(m,l)$ must satisfy $\Pr\Set{\RMX = \Mx }= f(\Mx|m,l)$. It
effectively adds additional noise to make it hard for Eve to learn the
secret. This artificial noise also affects Bob since he does no know
it a priori.

Alice then transmits the random codeword $\RMX$ through $n$ uses of
the wiretap channel. The \emph{$n$-th extension} of the wiretap
channel is characterized by the $n$-th direct power $(W_b^n,W_e^n)$,
where $W_b^n(\My|\Mx)=\prod_{i=1}^n W_b(y^{(i)}|x^{(i)})$ and
similarly for $W_e^n$. Bob uses his channel output $\RMY$ to decode
both the public and private messages with a deterministic decoder
$`f_b:\setY^n\mapsto M \times L$. $`F_b:M \times L \mapsto 2^{\setY^n}$
denotes the \emph{decision region} so that
\begin{align}
  `f_b(\My)&=(m,l)
  \iff \My \in `F_b(m,l) \label{eq:decision}
\end{align}
Similarly, Eve uses her channel output $\RMZ$ to decode the public
message with decoder $`f_e:\setZ^n \mapsto M$ and decision region
$`F_e:M \mapsto 2^{\setZ^n}$.  She, however, also generates an
unordered set of $`l\leq L$ distinct guesses of the secret using a
list decoder $`j: \setZ^n\mapsto \Set{\setA\subset L: \abs\setA=`l}$,
which is a \emph{correspondence}. The decision region $`J:L \mapsto
2^{\setZ^n}$ satisfies
\begin{align}
  l\in`j(\Mz)\iff \Mz \in `J(l)
\end{align}
The triple $(f,`f_b,`f_e)$ will be called an ($n$-block) \emph{wiretap
  channel code}, while the list decoder $`j$ will be called the
\emph{list decoding attack} (with deterministic list size). The quadruple
$(f,`f_b,`f_e,`j)$ will be called an ($n$-block) \emph{transmission
  (model)} for the wiretap channel.

\subsection{Achievable rate and exponent triples}
\label{sec:achi-rate-expon}

The performance of a wiretap channel code with respect to
a list decoding attack is evaluated based on the following fault events.

\begin{definition}[Fault events]
  \label{def:fault}
  Let $\setE_b(m,l)$, $\setE_e(m,l)$ and $\setS_e(m,l)$ be the fault
  events that Bob decodes $(m,l)$ wrong, Eve decodes $m$ wrong, and
  Eve successfully guesses $l$ respectively when $(m,l)$ is the public
  and private message pair. i.e.
  \begin{align*}
    \setE_b(m,l)&:=\Set*{`f_b(\RW_b^n(\Rf(m,l)))\neq (m,l)}\\
    \setE_e(m,l)&:=\Set*{`f_e(\RW_e^n(\Rf(m,l)))\neq m}\\
    \setS_e(m,l)&:=\Set*{l\in `j(\RW_e^n(\Rf(m,l)))}
  \end{align*}
  The corresponding \emph{(average) fault probabilities} (over the
  message set $M\times L$), $e_b$, $e_e$ and $s_e$ can be computed as
  follows.
  \begin{subequations}
    \label{eq:p}
    \begin{align}
      e_b
      &
      = \Avg_{m\in M,l\in L} \sum_{\Mx\in\setX^n} W_b^{n}(`F_b^c(m,l)|\Mx)f(\Mx|m,l)\label{eq:e_b}\\
      e_e
      &
      = \Avg_{m\in M,l\in L} \sum_{\Mx\in\setX^n} W_e^{n}(`F_e^c(m)|\Mx)f(\Mx|m,l)\label{eq:e_e}\\
      s_e
      &
      = \Avg_{m\in M,l\in L} \sum_{\Mx\in\setX^n} W_e^{n}(`J(l)|\Mx)f(\Mx|m,l)\label{eq:s_e}
    \end{align}
  \end{subequations}
  where $`F_b^c(m,l)$ and $`F_e^c(m)$ are the complements of the
  $`F_b(m,l)$ and $`F_e(m)$ respectively; and
  $\Avg_{m\in M, l\in L}$ denotes
  $\frac1{ML}\sum_{m\in M, l \in L}$. When there is ambiguity, we will
  write $e_b(f,`f_b,W_b)$ etc.\ to explicitly state its dependencies.
\end{definition}

We study the asymptotic properties when the sizes $M$ and $L$ of the
message sets and $`l$ of Eve's guessing list grow exponentially while
the fault probabilities decay exponentially in $n$. The exponential
rates are defined as follows.

\begin{definition}
  \label{def:rates}
  Consider a sequence of $n$-block transmissions
  $(f^{(n)},`f_b^{(n)},`f_e^{(n)},`j^{(n)})$ ($n\in`Z^+$) over the
  wiretap channel $(W_b,W_e)$, the \emph{public message rate} $R_M$,
  \emph{private message rate} $R_L$ and the \emph{guessing rate}
  $R_{`l}$ are defined as,
  \begin{subequations}
    \label{eq:R}
    \begin{align}
      R_M&:=\liminf_{n\to`8} \frac1n \log M^{(n)}\label{eq:R_M}\\
      R_L&:=\liminf_{n\to`8} \frac1n \log L^{(n)}\label{eq:R_L}\\
      R_{`l}&:=\limsup_{n\to`8} \frac1n \log `l^{(n)}\label{eq:R_`l}
    \end{align}
  \end{subequations}
  The exponents of the fault probabilities \eqref{eq:p} are defined
  as,
  \begin{subequations}
    \label{eq:E}
    \begin{align}
      E_b&:=\liminf_{n\to`8} -\frac1n \log e_b^{(n)}\label{eq:E_b}\\
      E_e&:=\liminf_{n\to`8} -\frac1n \log e_e^{(n)}\label{eq:E_e}\\
      S_e&:=\liminf_{n\to`8} -\frac1n \log s_e^{(n)}\label{eq:S_e}
    \end{align}
  \end{subequations}
  where $e_b^{(n)}$ and alike denotes $e_b$ evaluated with respect to
  the $n$-block transmission. For simplicity, the superscript $(n)$
  will be omitted hereafter if there is no ambiguity.
\end{definition}
In the \emph{code design phase} prior to the transmission phase, Alice
chooses $(f,`f_b,`f_e)$ without knowledge of $`j$ and then Eve chooses
$`j$ knowing Alice's choice. In particular, Eve chooses $`j$ to
minimize $S_e$ so that her success probability $s_e^{(n)}$ decays to
zero as slowly as possible, while Alice chooses $(f,`f_b,`f_e)$ to
make $E_b$, $E_e$ and $S_e$ large so that the error probabilities
$e_b^{(n)}$ and $e_e^{(n)}$ decay to zero fast for reliability, and
the probability $s_e^{(n)}$ of successful attack by Eve decays to zero
fast for secrecy. The \emph{tradeoff between secrecy and reliability}
for Alice can be expressed in terms of the set of achievable rate and
exponent triples defined as follows.

\begin{definition}[Achievable rate and exponent triples]
  \label{def:achievable}
  The rate triple $(R_1,R_2,R_3)\in `R_+^3$, where
  $`R_+:=\Set{a\in`R:a\geq 0}$, is \emph{achievable} if there exists a
  sequence of wiretap channel codes $(f,`f_b,`f_e)$ with rates,
  \begin{align*}
    R_M \geq R_1 \quad\text{and}\quad
    R_L \geq R_2
  \end{align*}
  such that for \emph{any sequence} of list decoding attack $`j$ with
  guessing rate $R_{`l} \leq R_3$, the probabilities $e_b$, $e_e$ and
  $s_e$ converge to zero as $n\to`8$. 

  The exponent triple $(E_1,E_2,E_3)\in `R_+^3$ is \emph{achievable}
  with respect to the rate triple if in addition that,
  \begin{align*}
    E_b \geq E_1 \quad\text{and}\quad
    E_e \geq E_2 \quad\text{and}\quad
    S_e \geq E_3
  \end{align*}
  If the achievable exponents are strictly positive, the rate triple
  is said to be \emph{strongly achievable}.
\end{definition}

In the sequel, we will obtain an inner bound to the set of achievable
exponent triples in the form of parameterized single-letter lower
bounds, one for each exponent.\footnote{In response to the question of
  using average instead of maximum error probabilities (over the
  message set), we would like to point out that the particular inner
  bound to be derived also holds when $e_b$ and $e_e$ are defined as
  the corresponding maximum error probabilities and $s_e$ as the
  average success probability. It follows from the usual argument of
  successively expurgating worst half of the codewords as in
  \cite{korner1980}, which turns out to preserve the desired
  \emph{overlap property} of the code and hence the bound for the
  success exponent. (see Section~\ref{sec:success-exponent}) If one
  defined $s_e$ as the maximum probability however, the problem
  becomes degenerate since there is an obvious strategy for Eve to
  achieve $s_e=1$.} From this, an inner bound to the set of strongly
achievable rate triples will be obtained, the closure of which
coincides with the closure of the achievable region in Theorem~1 of
\cite{csiszar1978} when the guessing rate is treated as equivocation
rate.

\section{Coding scheme}
\label{sec:coding-scheme}

The coding scheme (i.e.\ the specification of the sequence of wiretap
channel codes $(f,`f_b,`f_e)$, see \figref{fig:transmission})
considered here is a merge of the schemes in \cite{csiszar1978} and
\cite{korner1980} using the \emph{method of types} developed by
Csisz\'ar\cite{csiszar1981}. We will describe each key component of
the code in succession and explain how each of them simplifies the
analysis of the fault events (see \defref{def:fault}).

\subsection{Constant composition code}
\label{sec:const-comp-code}

As a first step, output of the stochastic encoder is restricted to
\emph{constant composition code}\cite{csiszar1981} defined as follows.
Let $N(x|\Mx)$ denote the number of occurrences of symbol $x\in\setX$
in the $n$-sequence $\Mx\in\setX^n$. The \emph{type} or
\emph{empirical distribution} $P_{\Mx}$ of $\Mx$ is defined as the probability
mass function,
\begin{align}
  P_{\Mx}(x)&:=\frac{N(x|\Mx)}{n} && \forall x\in\setX
  \label{eq:type}
\end{align}
Let $\rsfsP_n(\setX):=\Set{P_{\Mx}:\Mx\in\setX^n}$ denote the set of
all possible types of an $n$-sequence in $\setX^n$.
The type class $T_Q^{(n)}:=\Set{\Mx:P_{\Mx}=Q}$ or $T_Q$ for short
denotes the set of all $n$-sequences $\Mx$ having type $Q\in
\rsfsP_n(\setX)$. \emph{An $n$-block constant composition code $`q$
  on $\setX$ is an ordered tuple of codewords all from the same type
  class on $\setX$.} i.e.\ $\exists Q\in\rsfsP_n(\setX), `q\subset
T_Q$.

Suppose $`q$ is the constant composition code of type $Q$ for the
stochastic encoder $f$. Then, $f(\Mx|m,l)=0$ for all $\Mx\notin `q$. From
\eqref{eq:e_b},
\begin{align}
  e_b
  &= \Avg_{m\in M,l\in L} \sum_{\Mc\in`q}
  W_b^{n}(`F_b^c(m,l)|\Mc)f(\Mc|m,l) \label{eq:e_b1}
\end{align}
and similarly for other probabilities in \eqref{eq:p}. To further
simplify the expressions, define the \emph{canonical conditional type}
$P_{\My|\Mx}$ of $\My$ given $\Mx$ as,
\begin{align}
  P_{\My|\Mx}(y|x) &:= 
  \begin{dcases}
    1/\abs\setY & ,N(x|\Mx)=0\\
    \frac{N(x,y|\Mx,\My)}{N(x|\Mx)} & ,\text{otherwise}
  \end{dcases}
  \label{eq:canonical}
\end{align}
for all $x\in \setX, y\in\setY$, where $N(x,y|\Mx,\My)$ is the number
of occurrences of the pair $(x,y)$ in the $n$-sequence
$\Set{(x^{(i)},y^{(i)})}_{i=1}^n$ of pairs. The canonical conditional
type of $\My$ given $\Mx$ exists and is unique by
definition.\footnote{ This is a minor modification of the conditional
  type defined in Definition~1.2.4 of \cite{csiszar1981}, according to
  which $\My$ may have a continuum of conditional types $V$ given
  $\Mx$ since $V(y|x)$ can be arbitrary when $N(x|\Mx)=0$.}  However,
with a canonical conditional type $V$ given $\Mx$ specified, there can
be more than one $\My$ satisfying it.\footnote{For example, the binary
  sequences $1100$ and $0011$ have the same canonical conditional type
  given $1111$, i.e.\ $\bM .5 & .5\eM$.  Similarly, $1111$ has the
  same canonical conditional type whether it is given $1100$ or
  $0011$, i.e.\ $\bM 0 & 1\eM$.} If $V:\setX\mapsto\setY$ is the
conditional type of $\My$ given $\Mx$, $\My$ is said to lie in
$T_V(\Mx)$, referred to as the \emph{$V$-shell} of $\Mx$ or the
\emph{conditional type class} of $V$ given $\Mx$. In other words,
$T_V(\Mx)$ is the set of all $\My\in\setY^n$ with conditional type $V$
given $\Mx$.

Writing $W_b^n(\My|\Mc)$ as the product
$\prod_{x,y} W_b(y|x)^{N(x,y|\Mc,\My)}$,
Lemma~1.2.6 of \cite{csiszar1981} gives, for all $\My\in T_V(\Mc)$,
\begin{subequations}
\label{eq:c1}
\begin{align}
  W_b^n(\My|\Mc) &= \exp\Set*{-n[D(V\|W_b|Q)+H(V|Q)]}
  \label{eq:c1a}\\
  &= \frac{W_b^n(T_V(\Mc)|\Mc)}{\abs*{T_V(\Mc)}} \label{eq:c1b}
  \quad \because\text{\eqref{eq:c1a} is uniform}
\end{align}
\end{subequations}
where the \emph{conditional information divergence} $D(V\|W_b|Q)$ and
\emph{conditional entropy} $H(V|Q)$ are defined as,
\begin{align}
  D(V\|W|Q)&:= \sum_{(x,y)\in\setX\times\setY} Q(x)V(y|x) \ln
  \frac{V(y|x)}{W(y|x)} \label{eq:D}\\
  H(V|Q)&:=\sum_{(x,y)\in\setX\times\setY} Q(x)V(y|x)
  \ln\frac{1}{V(y|x)} \label{eq:H}
\end{align}
The key implication is that $W_b^n(\My|\Mc)$ depends on $\My$
only through the conditional type $P_{\My|\Mc}$ and channel output
$\RW_b^n(\Mc)$ is uniformly distributed within every $V$-shell $T_V(\Mc)$.

Let $\rsfsV_n(Q,\setY):=\Set{P_{\My|\Mx}:\Mx\in T_Q,\My\in\setY^n}$
($\rsfsV_n(Q)$ or $\rsfsV_n$ for short) be the set of all possible
canonical conditional types of $\My$ given $\Mc$. This set depends on
$\Mc$ only through the type $Q$ of $\Mc$.\footnote{For example, if
  $\My=011$ is in the $V$-shell of $\Mc=011$, then permutation
  $\My'=110$ of $\My$ is in the $V$-shell of the same permutation
  $\Mc'=110$ of $\Mc$. In general, if $V$ is a canonical type of some
  sequence $\My\in\setY^n$ given $\Mc\in`q$ then the $V$-shell of
  another codeword $\Mc'\in`q$ must contain a sequence
  $\My'\in\setY^n$, namely the sequence obtained from $\My$ by the
  same permutation of $\Mc\in`q$ to $\Mc'\in`q$. Thus, the set of all
  possible canonical conditional types are the same if the
  conditioning sequences have the same type.}
$\Set{T_V(\Mc):V\in\rsfsV_n(Q)}$ is a partitioning of $\setY^n$ for
every $\Mc\in`q$ because every $\My$ has a unique canonical
conditional type given $\Mc$. We can therefore partition the
probabilities by $\rsfsV_n(Q)$ as follows. From \eqref{eq:e_b1},
\iftwocolumn
\begin{align*}
    e_b
    &= \Avg_{m,l} \sum_{\Mc\in`q} \sum_{V\in\rsfsV_n} 
    W_b^{n}(`F_b^c(m,l)\cap T_V(\Mc)|\Mc)f(\Mc|m,l) 
    \notag\\
    &= \sum_{V\in\rsfsV_n} \sum_{\Mc\in`q}
    W_b^{n}(T_V(\Mc)|\Mc) \Avg_{m,l}
    \frac{\abs*{`F_b^c(m,l)\cap T_V(\Mc)}}{\abs*{T_V(\Mc)}}f(\Mc|m,l)
    \label{eq:e_b2}
\end{align*}
\else
\begin{align*}
    e_b
    &= \Avg_{m\in M,l\in L} \sum_{\Mc\in`q} \sum_{V\in\rsfsV_n(Q)} 
    W_b^{n}(`F_b^c(m,l)\cap T_V(\Mc)|\Mc)f(\Mc|m,l) 
    \notag\\
    &= \sum_{V\in\rsfsV_n(Q)} \sum_{\Mc\in`q}
    W_b^{n}(T_V(\Mc)|\Mc) \Avg_{m\in M,l\in L}
    \frac{\abs*{`F_b^c(m,l)\cap T_V(\Mc)}}{\abs*{T_V(\Mc)}}f(\Mc|m,l)
    \label{eq:e_b2}
\end{align*}
\fi
where the last equality is due to the piecewise uniform distribution
of the channel output $\RW_b^n(\Mc)$ implied by \eqref{eq:c1}.  By
Lemma~1.2.6 of \cite{csiszar1981},\footnote{The key step in the
  derivation is that $V^n(T_V(\Mc)|\Mc)\leq 1$ implies
  $\abs{T_V(\Mc)}\leq \exp\Set{nH(V|Q)}$ by \eqref{eq:c1} with $W_b$
  replaced by $V$.}
\begin{align}
  W_b^n(T_V(\Mc)|\Mc) \leq \exp\Set*{-n D(V\|W_b|Q)}
  \label{eq:c2}
\end{align}
Thus, $e_b$ can be upper bounded as,
\iftwocolumn
\begin{multline}
  e_b
  \leq \sum_{V\in\rsfsV_n(Q)} \exp\Set*{-nD(V\|W_b|Q)}\times\\\times
  \sum_{\Mc\in`q} 
  \Avg_{m\in M,l\in L}
  \frac{\abs*{`F_b^c(m,l)\cap T_V(\Mc)}}{\abs*{T_V(\Mc)}}f(\Mc|m,l)
  \label{eq:e_b3}
\end{multline}
\else
\begin{align}
  e_b
  \leq \sum_{V\in\rsfsV_n(Q)} \exp\Set*{-nD(V\|W_b|Q)}
  \sum_{\Mc\in`q} 
  \Avg_{m\in M,l\in L}
  \frac{\abs*{`F_b^c(m,l)\cap T_V(\Mc)}}{\abs*{T_V(\Mc)}}f(\Mc|m,l)
  \label{eq:e_b3}
\end{align}
\fi
\subsection{Transmission of junk data and prefix DMC}
\label{sec:transm-junk-data}

In the previous section, the use of constant composition code
simplifies the probability \eqref{eq:e_b} to \eqref{eq:e_b3} and
similarly for other probabilities in \eqref{eq:p}. In this section, we
shall specify the structure of the stochastic encoder $f$ and its
uniform randomization over junk data as follows.

Consider indexing the codewords in $`q$ as $\Mc_{jlm}$ by $j\in J$, $l\in
L$ and $m\in M$. i.e.\ 
\begin{align}
  \label{eq:ccc}
  `q&:=\Set{\Mc_{jlm}}_{j\in J,l\in L,m\in M}
\end{align}
Set $\Rf(m,l)= \Mc_{\RJ lm}$ where the junk data
$\RJ$ is a random variable Alice chooses uniformly randomly from
$\Set{1,\dots,J}$. The conditional probability $f$ is,
\begin{align}
  f(\Mc|m,l) 
  =
  \begin{cases}
    \frac1J & ,\text{if $\Mc\in\Set{\Mc_{jlm}:j\in J}$}\\
    0 &,\text{otherwise}
  \end{cases}\label{eq:junkdata}
\end{align}
This approach of providing secrecy, illustrated in
Example~\ref{eg:junkdata} in the Appendix, will be called
\emph{transmission of (uniformly random) junk data} because $\RJ$ is
not meant to be a message although it is encoded like one.\footnote{It
  turns out that $\RJ$ can also be reliably decoded by Bob with lower
  level of secrecy. Thus, one may choose $\RJ$ to be meaningful
  private data to achieve a new notion of \emph{unequal security
    protection}. However, it suffices for our case of interest to
  treat $\RJ$ as meaningless.} Substituting this into the upper bound
of $e_b$ in \eqref{eq:e_b3} and similarly for the other fault
probabilities gives the following expressions.

\begin{lem}[Constant composition code, transmission of junk data]
  \label{lem:ap}
  Using $n$-block constant composition code $`q$ in \eqref{eq:ccc} of
  type $Q\in\rsfsP(\setX)$ and the transmission of junk data
  approach \eqref{eq:junkdata}, the probabilities in \eqref{eq:p}
  can be upper bounded as follows,
  \iftwocolumn
  \begin{subequations}
    \label{eq:ap}
    \begin{align}
      e_b
      &\leq \smashoperator{\sum_{V\in\rsfsV_n(Q)}} 
      \exp\Set*{-nD(V\|W_b|Q)}
      \Avg_{j,l,m}
      \tfrac{\abs*{`F_b^c(m,l)\cap
          T_V(\Mc_{jlm})}}{\abs*{T_V(\Mc_{jlm})}}\label{eq:ae_b}\hspace*{-1em}\\
      e_e
      &\leq 
      \smashoperator{\sum_{V\in\rsfsV_n(Q)}} 
      \exp\Set*{-nD(V\|W_e|Q)}
      \Avg_{j,l,m}
      \tfrac{\abs*{`F_e^c(m)\cap
          T_V(\Mc_{jlm})}}{\abs*{T_V(\Mc_{jlm})}}\label{eq:ae_e}\hspace*{-1em}\\
      s_e
      &\leq 
      \smashoperator{\sum_{V\in\rsfsV_n(Q)}} 
      \exp\Set*{-nD(V\|W_e|Q)}
      \Avg_{j,l,m}
      \tfrac{\abs*{`J(l)\cap
          T_V(\Mc_{jlm})}}{\abs*{T_V(\Mc_{jlm})}}\label{eq:as_e}
    \end{align}
  \end{subequations}
  where $\Avg_{j,l,m}$ is over $j\in J$,
  $l\in L$ and $m\in M$.
  \else
  \begin{subequations}
    \label{eq:ap}
    \begin{align}
      e_b
      &\leq \sum_{V\in\rsfsV_n(Q)}
      \exp\Set*{-nD(V\|W_b|Q)}
      \Avg_{j\in J,l\in L,m\in M} 
      \frac{\abs*{`F_b^c(m,l)\cap
          T_V(\Mc_{jlm})}}{\abs*{T_V(\Mc_{jlm})}}\label{eq:ae_b}\\
      e_e
      &\leq 
      \sum_{V\in\rsfsV_n(Q)}
      \exp\Set*{-nD(V\|W_e|Q)}
      \Avg_{j\in J,l\in L,m\in M} 
      \frac{\abs*{`F_e^c(m)\cap
          T_V(\Mc_{jlm})}}{\abs*{T_V(\Mc_{jlm})}}\label{eq:ae_e}\\
      s_e
      &\leq 
      \sum_{V\in\rsfsV_n(Q)} 
      \exp\Set*{-nD(V\|W_e|Q)}
      \Avg_{j\in J,l\in L,m\in M} 
      \frac{\abs*{`J(l)\cap
          T_V(\Mc_{jlm})}}{\abs*{T_V(\Mc_{jlm})}}\label{eq:as_e}
    \end{align}
  \end{subequations}
  \fi
\end{lem}
Note that the randomization in the encoder is equivalent to the
averaging over the message augmented with junk data.

Another approach of randomization introduced in \cite{csiszar1978} is
the \emph{prefix discrete memoryless channel (prefix DMC)}, which is
characterized by the conditional probability distribution
$\tilde{V}\in\rsfsP(\setX)^{\tsetX}$ from some finite set
$\tsetX$. The stochastic encoder first maps $(m,l)$ into an
$n$-sequence in $\tsetX^n$, which is then fed through the extended prefix DMC
$\tilde{V}^n$ before being transmitted through the channel. To combine
this with the transmission of junk data approach, let $\tilde{f}$ be
the original stochastic encoder defined in \eqref{eq:junkdata} except
that $\setX$ is replaced by $\tsetX$, and $`q$ is a constant
composition code with type $Q$ on $\tsetX$. Then, the new encoder is,
\begin{align*}
  f(\Mx|m,l) 
  &:= \sum_{\Mc\in `q}{\tilde V}^n(\Mx|\Mc) \tilde{f}(\Mc|m,l)
  && \forall m\in M, l\in L, \Mx\in \setX^n
\end{align*}
This is illustrated in \figref{fig:encoder_orig}.

\begin{figure}
  \centering
  \input{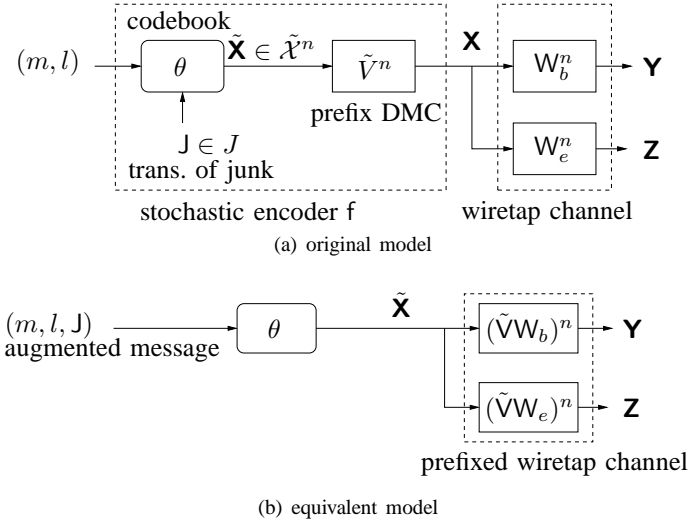}
  \caption{Stochastic encoding with transmission of junk data and prefix DMC}
  \label{fig:encoder}
\end{figure}

The prefix DMC can be viewed as part of the wiretap channel instead of
the encoder as in \figref{fig:encoder_eq} because the wiretap channel
$(W_b,W_e)$ prefixed with any discrete memoryless channel $\tilde V$
is just another wiretap channel $(\tilde V W_b,\tilde V W_e)$, where
the product $\tilde V W_b$ is the matrix multiplication.  Thus, any
performance metric, say $e(W_b,W_e)$, that one obtains without prefix
discrete memoryless channel can be converted to the performance metric
with prefixing discrete memoryless channel as $e(\tilde V W_b,\tilde V
W_e)$.

Because of this simplicity in extending any performance metrics with
prefix DMC, we will \emph{leave this prefixing procedure to the very
  end and use the encoder defined in \eqref{eq:junkdata} for the main
  analysis}. For a simple comparison
between the prefix DMC and transmission of junk data approach, readers
can refer to Example~\ref{eg:junkdata} and \ref{eg:prefixDMC}
in the Appendix.

\subsection{Random code construction and MMI decoding}
\label{sec:rand-code-constr}

As a summary, encoder $f$ encodes the public and private messages $m$
and respectively $l$, and the junk data $\RJ$ into a codeword
$\Mc_{\RJ lm}$ in the constant composition code $`q$ of type $Q$. The
codeword is then transmitted through the wiretap channel $(W_b,W_e)$,
to which a prefix a DMC $\Set{\tilde V}$ will be added in the end. The
fault probabilities simplify to \eqref{eq:ap}, with $(W_b,W_e)$
replaced by $(\tilde V W_b,\tilde V W_e)$ for the prefix DMC. It
remains to specify how the codebook $`q$ and decoders $(`f_b,`f_e)$
should be constructed.

Csisz\'ar and K\"orner\cite{csiszar1978} consider maximal code construction
with typical set decoding for the wiretap channel. This cannot be used
here since typical set decoding fails to give exponential decay rate
for the error probabilities. We will adopt the \emph{random code
construction scheme with maximum mutual information (MMI) decoding} in
\cite{korner1980} instead.


As a preliminary for the random code construction, some finite set
$\setU$ is chosen. The wiretap channel is \emph{trivially extended} with an
additional input symbol from $\setU$ to $(W_b\in \rsfsP(\setY)^{\setU\times
  \setX}, W_e\in\rsfsP(\setZ)^{\setU\times\setX})$, where
\begin{equation}
\label{eq:eW}
\begin{aligned}
  W_b(y|u,x)&:=W_b(y|x)\\
  W_e(y|u,x)&:=W_e(y|x)  
\end{aligned}
\end{equation}
for all $(u,x)\in\setU\times\setX$.  In the form of the stochastic
transition function, $\RW_b(u,x):=\RW_b(x)$ and $\RW_e(u,x):=\RW_e(x)$, which
means that the extended channel simply ignores the additional input
symbol. Thus, this trivial extension is purely conceptual and does not change
the original problem.


As the first step in the random code construction, a type
$Q_0\in\rsfsP_n(\setU)$ on $\setU$ is chosen for the constraint
length $n$. Then, each of the set $`Q_0:=\Set*{\RMU_m}_{m\in M}$ of
$n$-sequences is uniformly randomly and independently (u.i.) chosen from the
type class $T_{Q_0}$. i.e.
\begin{align*}
  P_{\RMU_m}(\Mu)
  &=
  \begin{cases}
    \frac1{\abs{T_{Q_0}}} & ,\Mu \in T_{Q_0}\\
    0 & ,\text{otherwise}
  \end{cases}
  && \forall m\in M
\end{align*}

Next, a conditional type $Q_1\;(\in\rsfsV_n(Q_0,\setX))$ is chosen. For each
$\RMU_m$ generated, consider its $Q_1$-shell $T_{Q_1}(\RMU_m)$. Each
of the set $`Q_1(m):=\Set{\RMX_{jlm}}_{j\in J,l\in L}$ of
$n$-sequences is chosen u.i.\ from
$T_{Q_1}(\RMU_m)$. i.e.
\begin{align*}
P_{\RMX_{jlm}|\RMU_m}(\Mx|\Mu) =
\begin{cases}
  \frac1{\abs{T_{Q_1}(\Mu)}} &,\Mx\in T_{Q_1}(\Mu)\\
  0 & ,\text{otherwise}
\end{cases}
\end{align*}
for all $(j,l,m)\in J\times L\times M,\Mu \in T_{Q_0}$.

Finally, $\RMU_m:=\Set{\RU_m^{(i)}}_{i=1}^n$ and
$\RMX_{jlm}:=\Set{\RX_{jlm}^{(i)}}_{i=1}^n$ are combined into one
codeword $\RMC_{jlm}:=\RMU_m\circ\RMX_{jlm}$, where $\circ$ denotes the
element-wise concatenation. i.e.\ 
\begin{align}
  \RMU_m\circ\RMX_{jlm}=\Set{(\RU_m^{(i)},\RX_{jlm}^{(i)})}_{i=1}^n
  \label{eq:circ}
\end{align}
The $i$-th term $\RC_{jlm}^{(i)}:=(\RU_m^{(i)},\RX_{jlm}^{(i)})$ is
transmitted in the $i$-th use of the (extended) wiretap channel.  The
random code $`Q$ is defined as the ordered structure
$\Set*{\RMC_{jlm}}_{j\in J,l\in L,m\in M}$. Its type is denoted as
$Q\in\rsfsP_n(\setU,\setX)$ where
$Q(u,x):=Q_0(u)Q_1(x|u)\;((u,x)\in\setU\times\setX)$. \emph{We write
  $Q=Q_0\circ Q_1$ where $\circ$ denotes the direct product.}

\begin{definition}[Random code]
  \label{def:random}
  The random code $`Q$ of type
  $Q:=Q_0\circ Q_1\;(Q_0\in \rsfsP_n(\setU),
  Q_1\in\rsfsV_n(Q_0,\setX))$ for the extended wiretap channel
  \eqref{eq:eW} is defined as follows,
  \begin{align*}
    `Q &:= \Set{\RMC_{jlm}}_{jlm}\\
    \RMC_{jlm} &:= \RMU_m \circ \RMX_{jlm}\\
    `Q_0 &:= \Set{\RMU_m}_m "<{^{\text{u.i.}}}-" T_{Q_0}\\
    `Q_1(m) &:= \Set{\RMX_{jlm}}_{jlm} "<{^{\text{u.i.}}}-" T_{Q_1}(\RMU_m)
  \end{align*}
  In words, it is the set of codewords $\RMC_{jlm}$ indexed by the
  messages $j\in J$, $l\in L$ and $m \in M$. Each codeword consists of
  an $n$-sequence $\RMU_m$ that belong to the random codebook $`Q_0$,
  and an $n$-sequence $\RMX_{jlm}$ that belongs to the random codebook
  $`Q_1(m)$. The codewords from $`Q_0$ are selected u.i.\ from the
  type class $T_{Q_0}$ and the codewords from $`Q_1(m)$ are selected
  u.i.\ from the $Q_1$-shell $T_{Q_1}(\RMU_m)$ of $\RMU_m$.
\end{definition}

This approach of random code construction is well-known in the
asymmetric broadcasting channel setting. $`Q_0$ is used to partition
$\setX^n$ into cells/clouds $\Set{T_{Q_1}(\RU_m)}_m$ that are intended
to be well distinguishable through the channels of both Bob and Eve,
and $`Q_1(m)$ are the set of codewords selected from the containing
cell that are intended to be well distinguishable by Bob but not
necessarily so by Eve. The addition of input symbol from $\setU$ gives
an additional degree of freedom in optimizing the average performance
of the code.

It is important to note that, unlike the randomness in the stochastic
encoding, the randomness in the codebook is known to all parties
(Alice, Bob and Eve). The randomization happens in the code design
phase before the public and private messages are generated for the
transmission phase.


With the structure of the codebook defined, we can now complete the
specification of the coding scheme with the \emph{maximum mutual
  information (MMI)} decoder for Bob and Eve. Consider a particular
realization $`q$ of the random code $`Q$. Let $I(Q,V)$ denote the
\emph{mutual information},
\begin{align}
  I(Q,V)&:=H(QV)-H(V|Q) && \text{see \eqref{eq:H}} \label{eq:I}
\end{align}
Then, $I(\Mc \wedge \My)$, referred to as the \emph{empirical mutual
  information} between $\Mx$ and $\My$, are defined as,
\begin{align}
  I(\Mx \wedge \My):= I(P_{\Mx},P_{\My|\Mx}) 
  && \text{see \eqref{eq:I},\eqref{eq:type},\eqref{eq:canonical}}
  \label{eq:eI}
\end{align}
Suppose Bob observes $\My\in\setY^n$ through his channel. He searches
for the codeword $\Mc\in`q$ that maximizes the empirical mutual
information $I(\Mc\wedge \My)$.\footnote{Note that the optimal
  decoding rule is the maximum likelihood decoding instead. MMI
  decoding is adopted here for simplicity.} If there is a unique
$\Mc_{jlm}$ that achieves the maximum, he declares $m$ as the public
message and $l$ as the private message. More precisely,
\iftwocolumn
\begin{multline}
  `f_b(\My) = (m,l) \iff \\
  \exists! (m,l,j),\, I(\Mc_{jlm} \wedge \My) = \max_{\Mc\in`q} I(\Mc\wedge
  \My)
  \label{eq:`f_b}
\end{multline}
\else
\begin{align}
  `f_b(\My) = (m,l) \iff 
  \exists! (m,l,j),\, I(\Mc_{jlm} \wedge \My) = \max_{\Mc\in`q} I(\Mc\wedge
  \My)
  \label{eq:`f_b}
\end{align}
\fi
Similarly, suppose Eve receives $\Mz$. She searches for the unique
$\Mu_m$ that achieves the maximum $\max_{\Mu \in`q_0} I(\Mu \wedge
\Mz)$.\footnote{One may think that Eve can
  search for the unique $\Mc_{jlm}$ that achieves the maximum
  $\max_{\Mc\in`q} I(\Mc \wedge \Mz)$, and declare $m$ as the public
  message. Because of the suboptimality of the MMI decoding and the
  random code construction, this choice turns out to be unfavorable.} i.e.
\begin{align}
  `f_e(\Mz) = m \iff
  \exists! m,\, I(\Mu_m \wedge \Mz) = \max_{\Mu \in `q_0}
  I(\Mu\wedge\Mz)
  \label{eq:`f_e}
\end{align}

We will not need to assume any structure for $`j$ other than the fact
it has to be a deterministic list decoder with fixed list size
$`l$.\footnote{It is clear, however, that the optimal $`j$ is an
  extension of the maximum likelihood decoding rule with $`l$ estimates
  instead of one.} The coding scheme without prefix DMC can now be
summarized as follows.

\begin{definition}[Coding scheme]
  \label{def:achieving}
  The coding scheme \emph{without} prefix DMC for
  a realization $`q$ of the random code in Definition~\ref{def:random}
  is defined as follows.
  \newline\noindent\textit{Encoding:}
  Alice generates the junk data $\RJ$ uniformly
    randomly from $\Set{1,\dots,J}$ and encodes the common message
    $m\in M$ and secret $l \in L$ into $(\Mu_m,\Mx_{\RJ
      lm})\in`q$. She only transmits $\RX_{\RJ lm}$ through the
    channel. The encoding function is therefore,
    \begin{align*}
      f(\Mx|m,l) &:=
      \begin{cases}
        \frac1J & \Mx \in \Set*{\Mx_{jlm}}_{j\in J}\\
        0 & ,\text{otherwise}
      \end{cases}\quad\text{, or equivalently}\\
      \Rf(m,l) &:= \Mx_{\RJ lm} \in `q_1(m) \quad ,\forall m\in M,l\in L
    \end{align*}
    \newline\noindent\textit{Decoding:}
    If Bob receives $\My$, he finds a codeword $\Mc\in`q$
    that maximizes the empirical mutual information $I(\Mc\wedge \My)$
    and use its location in $`q$ to decode $(m,l)$. The decoding
    function can be defined as,
    \begin{align*}
      `f_b(\My) = (m,l) \iff 
      \exists! (m,l,j),\, I(\Mc_{jlm} \wedge \My) = \max_{\Mc\in`q} I(\Mc\wedge \My)
    \end{align*}
    
    Similarly, Eve locates the mutual information maximizing codeword in
    $`q_0$ to decode $m$ as follows,
    \begin{align*}
      `f_e(\Mz) = m \iff
      \exists! m,\, I(\Mu_m \wedge \Mz) = \max_{\Mu \in `q_0}
      I(\Mu\wedge\Mz)
    \end{align*}
  The encoder and decoders are
  functions of the codebook $`q$, i.e.\ $f[`q](m,l|\Mc)$, $`f_b[`q](\My)$ and
  $`F_b[`q](\Mz;`q)$ etc.. However, for notational simplicity, the
  dependence on $`q$ will be omitted.
\end{definition}

Using the random coding scheme, we can further bound the fault
probabilities \eqref{eq:ap} with the expected fault probabilities over
the random code ensemble as follows.  From \eqref{eq:ae_b}, the
expectation of $e_b$ over the random code $`Q$ is,
\begin{align*}
  \opE(e_b(`Q))
  &\leq \sum_{V\in\rsfsV_n(Q)} \exp\Set*{-nD(V\|W_b|Q)}\times
  \\ &\qquad\times
  \overbrace{
  \opE`1(
  \Avg_{j\in J,l\in L,m\in M}
  \frac{\abs*{`F_b^c(m,l)\cap
      T_V(\RMC_{jlm})}}{\abs*{T_V(\RMC_{jlm})}}
  `2)}^{`b(V,`Q,`F_b^c):=}\\
  &\leq \abs{\rsfsV_n(Q)} \max_{V\in \setV_n(Q)}
  \exp\Set*{-nD(V\|W_b|Q)} `b(`Q,`F_b^c)\\
  &\hspace*{-1em}\leq (n+1)^{\abs\setX\abs\setY} \overbrace{\max_{V\in \rsfsV_n(Q)}
  \exp\Set*{-nD(V\|W_b|Q)} `b(`Q,`F_b^c)}^{s(W_b,`Q,`F_b^c):=}
\end{align*}
where the last inequality is due to the \emph{Type Counting Lemma}
$\abs{\rsfsV_n(Q)}\leq (n+1)^{\abs\setX\abs\setY}$.\footnote{ This
  follows from the definition \eqref{eq:canonical} that there are at
  most $n+1$ possible values for each entry of a canonical conditional
  type. (see Type Counting Lemma~2.2 of \cite{csiszar1981}.)} The
expectation of $e_e$ and $s_e$ can be upper bounded similarly. By the
union bound,
\begin{align*}
  \begin{multlined}[t]
    \Pr\big\{
    e_b(`Q) > 3 \opE(e_b(`Q)) \text{ or}\\
    e_e(`Q) > 3 \opE(e_e(`Q))\text{ or}\\
    s_e(`Q) > 3 \opE(s_e(`Q))
    \big\}
  \end{multlined}
  &\leq
  \begin{multlined}[t]
    \Pr\Set{e_b(`Q) > 3 \opE(e_b(`Q))}\\
    +\Pr\Set{e_e(`Q) > 3 \opE(e_e(`Q))}\\
    +\Pr\Set{s_e(`Q) > 3 \opE(s_e(`Q))}
  \end{multlined}
\end{align*}
which is $<1$ due to the Markov inequality $\Pr(\RA>`a
\opE(\RA))<1/`a$ for non-negative random variable $\RA$ and $`a>0$.
Thus, the complement of the event has positive probability, which
implies existence of a realization $`q$ of $`Q$ such that the fault
probabilities can be bounded simultaneously as follows,
\begin{subequations}
  \label{eq:apc}
  \begin{align}
    e_b(`q) &\leq 3(n+1)^{\abs\setX\abs\setY} s(W_b,`Q,`F_b^c)\label{eq:ae_bc}\\
    e_e(`q) &\leq 3(n+1)^{\abs\setX\abs\setY} s(W_e,`Q,`F_e^c)\label{eq:ae_ec}\\
    s_e(`q) &\leq 3(n+1)^{\abs\setX\abs\setY} s(W_e,`Q,`J)\label{eq:as_ec}
  \end{align}
\end{subequations}
where $s$ is defined as follows,
\begin{subequations}
  \label{eq:`bs}
  \begin{align}
    `b(V,`Q,`F)&:= \opE`1(\smashoperator[r]{\Avg_{j\in J,l\in L,m\in M}} \enspace\frac{\abs*{`F(m,l)\cap
        T_V(\RMC_{jlm})}}{\abs*{T_V(\RMC_{jlm})}}`2)\label{eq:`b}\hspace*{-.5em}\\
    s(W,`Q,`F)&:=\smashoperator{\max_{V\in\rsfsV_n(Q)}} \enspace\exp\Set*{-n D(V\|W|Q)}
    `b(V,`Q,`F)\label{eq:s}\hspace*{-.5em}
  \end{align}
\end{subequations}
and $`F_e(m,l):=`F_e(m)$, $`J(m,l):=`J(l)$ are the trivial
extensions for all $(m,l)\in M\times L$.

To compute the desired exponents, we consider a sequence of random
codes defined as follows.
\begin{definition}[Sequence of random codes]
  \label{def:randomseq}
  $\Set{`Q^{(n)}}$ or simply $`Q$ denotes a
  sequence of random codes $`Q^{(n)}$ (see
  Definition~\ref{def:random}) of type $Q^{(n)}=Q_0^{(n)}\circ
  Q_1^{(n)}\;(Q_0^{(n)}\in\rsfsP_n(\setU),
  Q_1^{(n)}\in\rsfsV_n(Q_0^{(n)},\setX))$ that converges to
  distribution $Q=Q_0\circ
  Q_1\;(Q_0\in\rsfsP(\setU),Q_1\in\rsfsP(\setX)^{\setU})$ in variation
  distance. i.e.\ $`d_{\op{var}}(Q^{(n)},Q)\to 0$, where
  \begin{align}
    `d_{\op{var}}(P,Q):=\max_{\setA \subset \setX} P(\setA) - Q(\setA)
    \label{eq:var}
  \end{align}
  Furthermore, $J^{(n)}$ grows exponentially at the junk data rate
  \begin{align}
    \lim_{n\to `8} \frac1n \log J^{(n)}=R_J\geq 0 \label{eq:R_J}
  \end{align}
\end{definition}

If one can find $`g_b(V,Q)$ continuous in $Q\circ V$ in variation
distance such that,
\begin{align*}
  \liminf_{n\to`8} - \frac1n \log `b(V^{(n)},`Q^{(n)},(`F_b^{(n)})^c) 
  &\geq `g_b(V,Q)
\end{align*}
for any $Q^{(n)}\circ V^{(n)}$ converging to $Q\circ V$ in variation distance,
then Bob's error exponent \eqref{eq:E_b} can be lower bounded as,
\begin{align*}
  E_b(`q) &=
  \liminf_{n\to`8} - \frac1n \log s(V^{(n)},`Q^{(n)},(`F_b^{(n)})^c) \\
  &\geq \min_{V\in\rsfsP(\setY)^{\setU\times\setX}} D(V\|W_b|Q)+`g_b(V,Q)
\end{align*}
and similarly for other exponents $E_e$ and $S_e$ in
\eqref{eq:E}.

\begin{lem}
  \label{lem:E}
  If $`g_b(V,Q)$, $`g_e(V,Q)$ and $`g(V,Q)$ are continuous in the
  joint distribution $Q\circ V$ (with respect to the variation
  distance \eqref{eq:var}) and lower bound the exponent,
  \begin{align*}
    \liminf_{n\to`8} - \frac1n \log `b(V,`Q,`F)
  \end{align*}
  for random code $`Q$ and the cases $`F$ equal to $`F_b^c$, $`F_e^c$ and $`J$
  respectively, then there exists a realization $`q$ of $`Q$ such that
  \begin{subequations}
    \label{eq:Ec}
    \begin{align}
      E_b(`q) &\geq \min_{V\in\rsfsP(\setY)^{\setU\times\setX}}
      D(V\|W_b|Q)+`g_b(V,Q)\label{eq:E_bc}\\
      E_e(`q) &\geq \min_{V\in\rsfsP(\setZ)^{\setU\times\setX}}
      D(V\|W_e|Q)+`g_e(V,Q)\label{eq:E_ec}\\
      S_e(`q) &\geq \min_{V\in\rsfsP(\setZ)^{\setU\times\setX}}
      D(V\|W_e|Q)+`g(V,Q)\label{eq:S_ec}
    \end{align}
  \end{subequations}
\end{lem}

In the sequel, we will compute $`g_b$, $`g_e$ and $`g$ to obtain the
desired lower bounds of the exponents.

\section{Success exponent}
\label{sec:success-exponent}

From Lemma~\ref{lem:E}, to obtain a lower bound of the
achievable\footnote{Achievable here does \emph{not} refer to
  achievable by Eve, but achievable by Alice as defined in
  Definition~\ref{def:achievable}.} success exponent $S_e$
\eqref{eq:S_e}, it suffices to compute a lower bound $`g(V)$ on the
exponent of the expected average fraction $`b(V,`Q,`J)$
for any $`J$ satisfying the guessing rate \eqref{eq:R_`l}.

\begin{figure}
  \centering
  \input{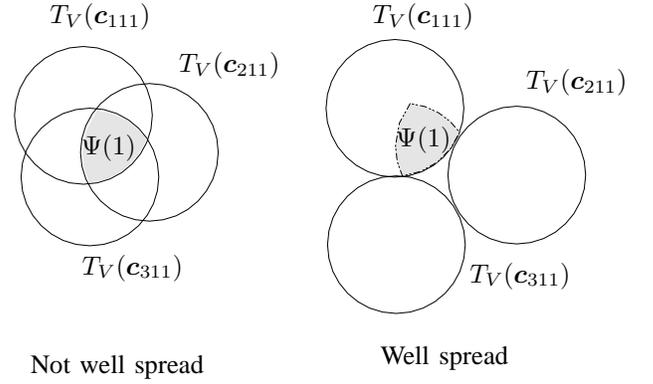}
  \caption{Effectiveness of stochastic encoding}
  \label{fig:hotspot}
\end{figure}

Consider first some realization $`q$ of the random code $`Q$ in
Definition~\ref{def:random}.
\begin{align*}
  `b(V,`q,`J) 
  &= \frac1{J\abs*{T_{V}(\Mc_{111})}} 
  \Avg_{l,m} 
  \sum_{j\in J} 
  \abs*{`J(l) \cap T_{V}(\Mc_{jlm})} && \text{by \eqref{eq:`b}}
\end{align*}
since $\abs*{T_{V}(\Mc_{jlm})}$ depends on $\Mc_{jlm}$ only through
its type $Q$ (and $n$).
The fraction can be made small if $\sum_j \abs*{`J(l) \cap
  T_{V}(\Mc_{jlm})}$ on the R.H.S.\ is made small for each $l$ and
$m$. Imagine $`J(l)$ as a net that Eve uses to cover the shells
$\Set{T_V(\Mc_{jlm}):j\in J}$ owned by Alice as much as possible.
Roughly speaking, since the net cannot be too large due to the list
size constraint, Alice should spread out the shells as much as
possible to minimize her loss. We will refer to this heuristically
desired property of $`q$ that the $V$-shells $\Set{T_V(\Mc_{jlm}):j\in
  J}$ spread out for every $V$, $m$ and $l$ as the \emph{overlap
  property}.\footnote{Though not explicitly stated, this notion of
  overlap property is also evident in \cite{csiszar1978} for the
  typical case when $V$ is close to $W_e$.  (See Lemma~2 of
  \cite{csiszar1978}) For the purpose of computing the exponent, we
  extend it to the atypical case of $V$ and relax the extent that the
  shells have to spread out by allowing subexponential amount of
  overlap.} This is illustrated in \figref{fig:hotspot}, in which the
configuration on the left has $\sum_{j=1}^3 \abs{`J(1)\cap
  T_V(\Mc_{j11})}$ three times larger than the one on the right.

Intuitively, random code has the \emph{overlap property} on average since
it uniformly spaces out the codewords. This is made precise with the
following \emph{Overlap Lemma}.

\begin{lem}[Overlap]
  Let $\RMX_j\;(j=1,\dots,J)$ be an $n$-sequence uniformly and
  independently drawn from $T_Q^{(n)}\subset \setX^n$. For all $J\in
  `Z^+$, $`d>0$, $n\geq n_0(`d,\abs\setX\abs\setZ)$, $\Mz\in \setZ^n$,
  $Q\in\rsfsP_n(\setX)$, $V\in\rsfsV_n(Q,\setZ)$ such that
  $\lfloor\exp\Set{nI(Q,V)}\rfloor\geq J$, we have,
  \begin{align*}
    \Pr\Set*{\sum_{j\in J} \ds1\Set{\Mz\in T_V(\RMX_j)}\geq \exp(n`d)}
    \leq \exp(-\exp(n`d))
  \end{align*}
  where $\ds1$ is the indicator function and $n_0$ is some
  integer-valued function that depends only on $`d$ and
  $\abs\setX\abs\setZ$.
\end{lem}

In words, the lemma states that the chance of having exponentially
($\exp(n`d)$) many shells (from $\Set{T_V(\RMX_j):j\in J}$)
overlapping at a spot ($\Mz$) is doubly exponentially decaying
($\exp(-\exp(n`d))$), provided that the shells are not enough to fill
the entire space ($T_{QV}\subset \setZ^n$) they can possibly reside.
(i.e.\ $J\leq \lfloor\exp\Set{nI(Q,V)}\rfloor$) For the case of
interest, we will prove the following more general form of the lemma
with conditioning.

\begin{lem}[Overlap (with conditioning)]
  \label{lem:overlap}
  Let $Q:=Q_0\circ Q_1\;(Q_0\in \rsfsP_n(\setU),
  Q_1\in\rsfsV_n(Q_0,\setX))$ be a joint type, $\RMU$ be a random
  variable distributed over $T_{Q_0}$, and $\RMX_j\;(j=1,\dots,J)$ be
  an $n$-sequence uniformly and independently drawn from
  $T_{Q_1}(\RMU)\subset \setX^n$. For all $J\in `Z^+$, $`d>0$, $n\geq
  n_0(`d,\abs\setU\abs\setX)$, $\Mz\in \setZ^n$, $Q:=Q_0\circ Q_1$,
  $V\in\setV_n(Q,\setZ)$ such that
  $\lfloor\exp\Set{nI(Q_1,V|Q_0)}\rfloor\geq J$, we have, \iftwocolumn
  \begin{multline}
    \Pr\Set*{\sum_{j\in J} \ds1\Set*{\Mz\in
        T_V(\RMU\circ \RMX_j)}\geq \exp(n `d)} \\
    \leq  \exp\Set{-\exp(n`d)} \label{eq:overlap}
  \end{multline}
  \else
  \begin{align}
    \Pr\Set*{\sum_{j\in J} \ds1\Set*{\Mz\in
        T_V(\RMU\circ \RMX_j)}\geq \exp(n `d)} 
    \leq  \exp\Set{-\exp(n`d)} \label{eq:overlap}
  \end{align}
  \fi
  where $\circ$ denotes element-wise concatenation~\eqref{eq:circ}, and
  \begin{equation}
    \label{eq:cI}
    \begin{aligned}
      I(Q_1,V|Q_0)&:=H(Q_1|Q_0)-H(V|Q_0\circ Q_1)\\
      &=H(Q_1|Q_0)-H(V|Q)
    \end{aligned}
  \end{equation}
  denotes the \emph{conditional mutual information}. (cf.\ \eqref{eq:I})
\end{lem}

\begin{proof}
  For notational simplicity, consider the case when $\exp(n`d)$ and
  $\exp\Set{nI(Q_1,V|Q_0)}$ are integers.\footnote{The case when
    $\exp(n`d)$ and $I(Q_1,V|Q_0)$ are not integers can be derived by
    taking their ceilings or floors and grouping the fractional
    increments into some dominating terms.} Consider some subset
  $\setJ$ of $\Set{1,\dots,J}$ with $\abs\setJ=\exp(n`d)$. Since the
  events $\Mz\in T_V(\RMU\circ\RMX_j)\;(j=1,\dots,J)$ are
  conditionally mutually independent given $\RMU=\Mu\in T_{Q_0}$,
  \iftwocolumn
  \begin{multline*}
    \Pr\Set*{\Mz\in \bigcap\nolimits_{j\in \setJ}
      T_V(\RMU\circ\RMX_j)}\\
    \begin{aligned}
      &= \sum_{\Mu\in T_{Q_0}} P_{\RMU}(\Mu) \Pr\Set{\Mz\in
        T_V(\Mu\circ\RMX_j)}^{\exp(n`d)}\\
      &\leq \exp\Set*{-n[I(Q_1,V|Q_0)-\frac{`d}2]\exp(n`d)}
    \end{aligned}
  \end{multline*}
  \else
  \begin{align*}
    \Pr\Set*{\Mz\in \bigcap\nolimits_{j\in \setJ}
      T_V(\RMU\circ\RMX_j)}
    &= \sum_{\Mu\in T_{Q_0}} P_{\RMU}(\Mu) \Pr\Set{\Mz\in
      T_V(\Mu\circ\RMX_j)}^{\exp(n`d)}\\
    &\leq \exp\Set*{-n[I(Q_1,V|Q_0)-\frac{`d}2]\exp(n`d)}
  \end{align*}
  \fi
  for $n\geq n'_0(`d,\abs\setU\abs\setX)$, where the last inequality
  is by Lemma~\ref{lem:rcwd} using the uniform distribution of
  $\RMX_j$ and Lemma~1.2.5 of \cite{csiszar1981} on the cardinality
  bounds of conditional type class. Since
  $\exp\Set{nI(Q_1,V|Q_0)}\geq J$, the number of distinct choices of
  $\setJ$ is,
  \begin{align*}
    {J \choose \exp(n`d)} &\leq {\exp(nI(Q_1,V|Q_0)) \choose
      \exp(n`d)}\\
    &\leq \exp\Set*{[\log e + n(I(Q_1,V|Q_0)-`d)]\exp(n`d)}
  \end{align*}
  where the last inequality is by Lemma~\ref{lem:choose}. By the union
  bound, L.H.S.\ of \eqref{eq:overlap} is upper bounded by the product
  of the last two expressions, i.e.
  \begin{align*}
    {J \choose \exp(n`d)} \Pr\Set*{\Mz\in \bigcap_{j\in \setJ}
      T_V(\RMU\circ\RMX_j)}
  \end{align*}
  Substituting the previously derived bounds for each term gives the
  desired upper bound $\exp(-\exp(n`d))$ when $n\geq
  n_0(`d,\abs\setU\abs\setX)$.
\end{proof}

Consider now a sequence of random codes $`Q^{(n)}$ defined in
Definition~\ref{def:randomseq}. The desired bound on the exponent of
$`b(V,`Q,`J)$ can be computed as follows using the Overlap Lemma.

\begin{lem}[Success exponent]
  \label{lem:woverlap}
  Consider the random code sequence $`Q$ defined in
  Definition~\ref{def:randomseq}. For any sequence of list decoding
  attack $`j$ satisfying the guessing rate $R_{`l}$ \eqref{eq:R_`l},
  \iftwocolumn
  \begin{multline*}
    \liminf_{n\to`8} -\frac1n \log `b(V,`Q,`J)\\
    \geq \abs{R_L - R_{`l} + \abs{R_J - I(Q_1,V|Q_0)}^-}^+
  \end{multline*}
  \else
  \begin{align*}
    \liminf_{n\to`8} -\frac1n \log `b(V,`Q,`J))
    \geq \abs{R_L - R_{`l} + \abs{R_J - I(Q_1,V|Q_0)}^-}^+
  \end{align*}
  \fi
  where $\abs{a}^+:=\max\Set{0,a}$ and $\abs{a}^-:=\min\Set{0,a}$.
\end{lem}

\begin{proof}
  By the Overlap Lemma~\ref{lem:overlap}, for any $`d>0$ and $n\geq
  n_0(`d)$,
  \iftwocolumn
  \begin{multline*}
    \Pr\Set*{\extendvert{\sum_{j\in \setJ_k(V)} \ds1\Set*{\Mz\in
        T_V(\RMC_{jlm})}\geq \exp(n `d) | `Q_0=`q_0}}\\
    \leq  \exp\Set{-\exp(n`d)}
  \end{multline*}
  \else
  \begin{align*}
    \Pr\Set*{\extendvert{\sum_{j\in \setJ_k(V)} \ds1\Set*{\Mz\in
        T_V(\RMC_{jlm})}\geq \exp(n `d) | `Q_0=`q_0}}
    \leq  \exp\Set{-\exp(n`d)}
  \end{align*}
  \fi
  where $`Q_0$ is the codebook $\Set{\RMU_m}_{m\in M}$, $`q_0$ is
  an arbitrary realization, and
  $\Set{\setJ_k(V)}_{k\in K_V}$ is a partitioning of
  $\Set*{1,\dots,J}$ defined as,
  \begin{align*}
    \setJ_k(V) &:= \Set*{(k-1)J_V+1,\dots,\min\Set{k J_V,J}}\\
    J_V &:= `1\lfloor \exp\Set{nI(Q_1,V|Q_0)}`2\rfloor\\
    K_V &:= `1\lceil J/J_V `2\rceil
  \end{align*}
  The expectation of the sum of indicators on the left can then be
  bounded as follows,
  \iftwocolumn
  \begin{multline*}
    \opE`1(\extendvert{\sum\nolimits_{j\in \setJ_k(V)} \ds1\Set*{\Mz\in
        T_V(\RMC_{jlm})} | `Q_0=`q_0} `2)\\
    \begin{aligned}
      &\leq \exp(n`d)\cdot 1 + J \cdot \exp\Set{-\exp(n`d)}\\
      &\leq \exp(n2`d)
    \end{aligned}
  \end{multline*}
  \else
  \begin{align*}
    \opE`1(\extendvert{\sum\nolimits_{j\in \setJ_k(V)} \ds1\Set*{\Mz\in
        T_V(\RMC_{jlm})} | `Q_0=`q_0} `2)
    &\leq \exp(n`d)\cdot 1 + J \cdot \exp\Set{-\exp(n`d)}\\
    &\leq \exp(n2`d)
  \end{align*}
  \fi
  where the last inequality is true for $n\geq
  n_0(`d,R_J,\abs\setU\abs\setX)$ by \eqref{eq:R_J}. Since
  $T_V(\RMC_{jlm})$ is contained by $T_{Q_1V}(\RMU_m)$,
  \begin{align*}
    &\sum_{\Mz\in `J(l)} \opE`1(\extendvert{\sum\nolimits_{j\in \setJ_k(V)} \ds1\Set*{\Mz\in
        T_V(\RMC_{jlm})} | `Q_0=`q_0 }`2)\\
    &= \sum_{\Mz\in `J(l)\cap T_{Q_1 V}(\Mu_m)} 
    \opE`1(\extendvert{\sum_{j\in \setJ_k(V)} \ds1\Set*{\Mz\in
        T_V(\RMC_{jlm})} | `Q_0=`q_0} `2)\\
    &\leq \exp(n2`d) \abs*{`J(l)\cap T_{Q_1 V}(\Mu_m)}
  \end{align*}
  By linearity of expectation,
  \iftwocolumn
  \begin{multline*}
    \opE`1(
    \extendvert{\sum\nolimits_{j\in \setJ_k(V)} \abs*{
        `J(l)\cap T_V(\RMC_{jlm})} | `Q_0=`q_0}
    `2)\\
    \leq \exp(n2`d) \abs*{`J(l)\cap T_{Q_1 V}(\Mu_m)}
  \end{multline*}
  \else
  \begin{align*}
    \opE`1(
    \extendvert{\sum\nolimits_{j\in \setJ_k(V)} \abs*{
        `J(l)\cap T_V(\RMC_{jlm})} | `Q_0=`q_0}
    `2)
    \leq \exp(n2`d) \abs*{`J(l)\cap T_{Q_1 V}(\Mu_m)}
  \end{align*}
  \fi
  Summing both sides over $k\in K_V$,
  \iftwocolumn
  \begin{multline*}
    \opE`1(
    \extendvert{\sum\nolimits_{j\in J} \abs*{
        `J(l)\cap T_V(\RMC_{jlm})} | `Q_0=`q_0}
    `2)\\
    \leq \exp(n2`d) K_V \abs*{`J(l)\cap T_{Q_1 V}(\Mu_m)}
  \end{multline*}
  \else
  \begin{align*}
    \opE`1(
    \extendvert{\sum\nolimits_{j\in J} \abs*{
        `J(l)\cap T_V(\RMC_{jlm})} | `Q_0=`q_0}
    `2)
    \leq \exp(n2`d) K_V \abs*{`J(l)\cap T_{Q_1 V}(\Mu_m)}
  \end{align*}
  \fi
  Summing both sides over $l\in L$ and applying the list size
  constraint on $`J$ in Lemma~\ref{lem:list} to the R.H.S.,
  \iftwocolumn
  \begin{multline*}
    \opE`1(
    \extendvert{\sum\nolimits_{j\in J,l\in L} \abs*{
        `J(l)\cap T_V(\RMC_{jlm})} | `Q_0=`q_0}
    `2) \\
    \leq \exp(n2`d) K_V `l \abs*{T_{Q_1 V}(\Mu_m)}
  \end{multline*}
  \else
  \begin{align*}
    \opE`1(
    \extendvert{\sum\nolimits_{j\in J,l\in L} \abs*{
        `J(l)\cap T_V(\RMC_{jlm})} | `Q_0=`q_0}
    `2) 
    \leq \exp(n2`d) K_V `l \abs*{T_{Q_1 V}(\Mu_m)}
  \end{align*}
  \fi
  Averaging both sides over $m\in M$, dividing by the
  constant $JL\abs*{T_V(\RMC_{jlm})}$ and taking the expectation over
  all possible realizations of $`q_0$ gives,
  \begin{align*}
    `b(V,`Q,`J)
    & \leq \exp(n2`d) \frac{K_V `l}{JL} \frac{\abs*{T_{Q_1
          V}(\Mu_1)}}{\abs*{T_{V}(\Mc_{111})}}
  \end{align*}
  To compute the desired exponent from the last inequality,
  denote the inequality in the exponent $\dotleq$ as follows,
  \begin{align}
    a_n \dotleq b_n  \iff
    \limsup_{n\to`8} \frac1n \log a_n \leq \liminf_{n\to`8} \frac1n
    \log b_n
  \end{align}
  Then, $K_V\dotleq \exp\Set{n\abs{R_J-I(Q_1,V|Q_0)}^+}$, $J\dotleq \exp\Set{nR_J}$ by \eqref{eq:R_J}, $L\dotgeq
  \exp\Set{nR_L}$ by \eqref{eq:R_L}, $`l\dotleq \exp\Set{nR_{`l}}$ by
  \eqref{eq:R_`l}, and $\abs*{T_{Q_1
        V}(\Mu_1)}/\abs*{T_{V}(\Mc_{111})}$ is $\dotleq
  \exp\Set{nI(Q_1,V|Q_0)}$. 
  Combining these, $`b(V,`Q,`J) $ is $\dotleq$ the following
  expression,
  \iftwocolumn
  \begin{multline*}
    \exp\Set{n[R_L-R_{`l}+[R_J-I(Q,V|Q_0)]\\
      -\abs{R_J-I(Q,V|Q_0)}^+]}
  \end{multline*}
  \else
  \begin{align*}
    \exp\Set{n[R_L-R_{`l}+[R_J-I(Q,V|Q_0)]-\abs{R_J-I(Q,V|Q_0)}^+]}
  \end{align*}
  \fi
  To obtain the desired bound, simplify this with the identity
  $\abs{a}^-\equiv a-\abs{a}^+$, and the fact that $`b(V,`Q,`J)\leq
  1$.
\end{proof}

\section{Error exponents}
\label{sec:error-exponents}

The desired error exponents can be obtained directly from the
achievability result in \cite{korner1980} by grouping $(j,l)\in
J\times L$ as one private message for Bob. This is because the error
exponent that Bob decodes the private message wrong lower bounds the
exponent that Bob decodes the secret wrong.\footnote{Since Bob can
  also decode the junk data as reliably as the secret, one may
  potentially transmit meaningful data instead of the junk provided
  that the data is uniformly random and need not be secured at the
  same level as the secret.}  For completeness, we provide a similar
derivation in this section. Readers familiar with \cite{korner1980}
and may skip to the next section.

In essence of Lemma~\ref{lem:E}, the error exponents for Bob and Eve
can be obtained by lower bounding the exponents of the fractions
$`b(V,`Q,`F_b^c)$ and respectively $`b(V,`Q,`F_e^c)$. Thus, the
objective is to prove the following lemma.

\begin{lem}[Error exponents]
  \label{lem:wpack}
  Consider the sequence of random code $`Q$ in
  Definition~\ref{def:randomseq}, and the MMI decoder
  (decision region map) $`f_b$ ($`F_b$) \eqref{eq:`f_b} and $`f_e$
  ($`F_e$) \eqref{eq:`f_e} for Bob and respectively Eve. Then,
  \iftwocolumn
  \begin{align*}
    &\liminf_{n\to`8} -\frac1n \log `b(V,`Q,`F_b^c))\\
    &\enspace\leq \abs*{I(Q_1,V|Q_0)-R_J-R_L + \abs*{I(Q_0,Q_1V)-R_M}^-}^+\\
    &\liminf_{n\to`8} -\frac1n \log `b(V,`Q,`F_e^c))
    \leq \abs*{I(Q_0,Q_1V)-R_M}^+
  \end{align*}
  \else
  \begin{align*}
    \liminf_{n\to`8} -\frac1n \log `b(V,`Q,`F_b^c))
    &\leq \abs*{I(Q_1,V|Q_0)-R_J-R_L + \abs*{I(Q_0,Q_1V)-R_M}^-}^+\\
    \liminf_{n\to`8} -\frac1n \log `b(V,`Q,`F_e^c))
    &\leq \abs*{I(Q_0,Q_1V)-R_M}^+
  \end{align*}
  \fi
\end{lem}

\subsection{Exponent for Bob}
\label{sec:exponent-bob}
In essence of Lemma~\ref{lem:E}, the error exponent for Bob can be
obtained by lower bounding the exponent of the fraction,
\begin{align*}
  `b(V,`Q,`F_b^c)
  &= \opE`1(
  \Avg_{j\in J, l\in L, m\in M} 
  \frac{\abs*{`F_b^c(m,l) \cap T_V(\RMC_{jlm})}}
  {\abs*{T_V(\RMC_{jlm})}}
  `2)
\end{align*}
where $`Q$ is the sequence of random codes in
Definition~\ref{def:randomseq} and $`F_b$ is the decision region of
the MMI decoder $`f_b$ in \eqref{eq:`f_b}.
$`F_b^c(m,l) \cap T_V(\RMC_{jlm})$ is the set of bad observations in
the $V$-shell of $\RMC_{jlm}$ that lead to error if $\RMC_{jlm}$ is
transmitted. With the MMI decoder \eqref{eq:`f_b}, this corresponds
to the set of $\My\in T_V(\RMC_{jlm})$ that has $I(\RMC_{jlm} \wedge
\My)$ no larger than $I(\RMC_{j'l'm'} \wedge \My)$ for some misleading
codeword $\RMC_{j'l'm'}$ where $j'\in J$ and $(l',m')\in L\times M
`/\Set{l,m}$. i.e.
\begin{multline*}
  `F_b^c(m,l) \cap T_V(\RMC_{jlm})
  = \{\My\in T_V(\RMC_{jlm}) \cap T_{V'}(\RMC_{j'l'm'}) :\\
  (j',l',m')\in \setW_b^{(1)}(m)\cup \setW_b^{(2)}(m,l),
  V'\in \setV_b(V)\}
\end{multline*}
where
\begin{align*}
  \setV_b(V)&:=\Set{V'\in \rsfsV_n(Q):I(Q,V')\geq I(Q,V)}\\
  \setW_b^{(1)}(m)
  &:=\Set{(j',l',m'):j'\in J, l'\in L, m' \in M`/\Set{m}}\\
  \setW_b^{(2)}(m,l)&:=\Set{(j',l',m):j'\in J, l'\in L`/\Set{l}}
\end{align*}
(The dependence on $V$, $m$ and $l$ will be omitted if there is no
ambiguity.) $\setV_b$ is the set of \emph{problematic conditional
  type} that can lead to error. $(\setW_b^{(1)},\setW_b^{(2)})$ forms
a partition of the set of \emph{indices for the misleading codewords}.
In particular, $\setW_b^{(1)}$ corresponds to the indices of
misleading codewords that result in decoding the public message wrong
if the observation lies in a problematic $V'$-shell of the misleading
codeword.  Similarly, $\setW_b^{(2)}$ corresponds to the indices of
misleading codewords that result in decoding the private message wrong
but decoding the public message correctly.\footnote{The reason for
  this separation is that the two types of error lead to two
  different exponents.}
By the union bound,
\iftwocolumn
\begin{multline*}
  \abs*{`F_b^c(m,l) \cap T_V(\RMC_{jlm})}\\
\begin{aligned}
  &\leq \sum_{V'\in \setV_b} \sum_{(j'l'm')\in W_b^{(1)}}
  \abs*{T_V(\RMC_{jlm}) \cap T_{V'}(\RMC_{j'l'm'})}
  \\ &\quad + \sum_{V'\in \setV_b} \sum_{(j'l'm')\in W_b^{(2)}}
  \abs*{T_V(\RMC_{jlm}) \cap T_{V'}(\RMC_{j'l'm'})}
\end{aligned}
\end{multline*}
\else
\begin{align*}
  \abs*{`F_b^c(m,l) \cap T_V(\RMC_{jlm})}
  &\leq \sum_{V'\in \setV_b} \sum_{(j'l'm')\in W_b^{(1)}}
  \abs*{T_V(\RMC_{jlm}) \cap T_{V'}(\RMC_{j'l'm'})}
  \\ &\quad + \sum_{V'\in \setV_b} \sum_{(j'l'm')\in W_b^{(2)}}
  \abs*{T_V(\RMC_{jlm}) \cap T_{V'}(\RMC_{j'l'm'})}
\end{align*}
\fi
Consider the second summation where $\RMU_{m'}=\RMU_{m}$ because
$m'=m$. Since $T_V(\RMC_{jlm}) \cap T_{V'}(\RMC_{j'l'm'})$ is
contained by
\begin{align*}
  T_{Q_1V}(\RMU_m)\cap T_{Q_1V'}(\RMU_m)
  \subset
  T_{QV}\cap T_{QV'}
\end{align*}
the summand is zero if $QV\neq QV'$ or $Q_1 V\neq Q_1 V'$ by the
uniqueness of (canonical conditional) types. Since the premise implies
$I(Q_0,Q_1V)=I(Q_0,Q_1V')$, we can impose this constraint
(temporarily) in the second summation without affecting the sum. Under
this equality constraint, however, the inequality constraint
$I(V|Q)\geq I(V'|Q)$ on $V'$ can be replaced by $I(Q_1,V|Q_0)\geq
I(Q_1,V'|Q_0)$. Withdrawing the equality constraint gives the
following upper bound,
\iftwocolumn
\begin{multline*}
  \abs*{`F_b^c(m,l) \cap T_V(\RMC_{jlm})}\\
  \leq\sum_{\substack{V'\in \rsfsV(Q):\\I(Q,V)\geq I(Q,V')}} 
  \smashoperator[r]{\sum_{(j'l'm')\in W_b^{(1)}}}
  \abs*{T_V(\RMC_{jlm}) \cap T_{V'}(\RMC_{j'l'm'})}\\
  + \hspace*{-1em}\sum_{\substack{V'\in\rsfsV(Q):I(Q_1,V|Q_0)\\\geq
      I(Q_1,V'|Q_0)}} 
  \smashoperator[r]{\sum_{(j'l'm')\in W_b^{(2)}}}
  \abs*{T_V(\RMC_{jlm}) \cap T_{V'}(\RMC_{j'l'm'})}
\end{multline*}
\else
\begin{align*}
  \abs*{`F_b^c(m,l) \cap T_V(\RMC_{jlm})}
  &\leq\sum_{\substack{V'\in \rsfsV(Q):\\I(Q,V)\geq I(Q,V')}} 
  \smashoperator[r]{\sum_{(j'l'm')\in W_b^{(1)}}}
  \abs*{T_V(\RMC_{jlm}) \cap T_{V'}(\RMC_{j'l'm'})}\\
  &\quad + \hspace*{-1em}\sum_{\substack{V'\in\rsfsV(Q):I(Q_1,V|Q_0)\\\geq
      I(Q_1,V'|Q_0)}} 
  \smashoperator[r]{\sum_{(j'l'm')\in W_b^{(2)}}}
  \abs*{T_V(\RMC_{jlm}) \cap T_{V'}(\RMC_{j'l'm'})}
\end{align*}
\fi

To bound the expectation on the left, it suffices to bound the
expectation of $\abs{T_V(\RMC_{jlm}) \cap T_{V'}(\RMC_{j'l'm'})}$ on
the right by the Packing Lemma\cite{csiszar1981}, which is stated
in a convenient form with conditioning in Lemma~\ref{lem:packing}.

If $(j',l',m')\in \setW_b^{(1)}(m)$, then $\RMC_{jlm}$ is independent
of $\RMC_{j'l'm'}$. Applying the Packing Lemma without conditioning
gives, for all $`d>0$, $n>n_0(`d,\abs\setU\abs\setX)$,
\iftwocolumn
\begin{align*}
  \opE`1(
  \tfrac{\abs{T_V(\RMC_{jlm}) \cap T_{V'}(\RMC_{j'l'm'})}}
  {\abs{T_V(\RMC_{jlm})}}`2)
  &\leq \exp\Set*{-n[I(Q,V')-`d]}
\end{align*}
\else
\begin{align*}
  \opE`1(
  \frac{\abs{T_V(\RMC_{jlm}) \cap T_{V'}(\RMC_{j'l'm'})}}
  {\abs{T_V(\RMC_{jlm})}}`2)
  &\leq \exp\Set*{-n[I(Q,V')-`d]}
\end{align*}
\fi
If $(j',l',m')\in \setW_b^{(2)}(m,l)$ instead, then $\RMC_{jlm}$ is
conditionally independent of $\RMC_{j'l'm'}$ given $\RMU_m$. The
Packing Lemma gives, 
\iftwocolumn
\begin{align*}
   \opE`1(
  \tfrac{\abs{T_V(\RMC_{jlm}) \cap T_{V'}(\RMC_{j'l'm'})}}
  {\abs{T_V(\RMC_{jlm})}}`2)
  &\leq \exp\Set*{-n[I(Q_1,V'|Q_0)-`d]}
\end{align*}
\else
\begin{align*}
   \opE`1(
  \frac{\abs{T_V(\RMC_{jlm}) \cap T_{V'}(\RMC_{j'l'm'})}}
  {\abs{T_V(\RMC_{jlm})}}`2)
  &\leq \exp\Set*{-n[I(Q_1,V'|Q_0)-`d]}
\end{align*}
\fi
Combining the last three inequalities, we have for $n$ sufficiently
large that,
\begin{align*}
  \opE`1(\frac{\abs*{`F_b^c(m,l) \cap
      T_V(\RMC_{jlm})}}{\abs*{T_V(\RMC_{jlm})}}`2)
  &\leq JLM \exp\Set{-n[I(Q,V)-`d]} 
  \\ & \hspace*{-1em}
  + JL\exp\Set*{-n[I(Q_1,V|Q_0)-`d]}
\end{align*}
where we have used the fact that $\abs{\setW^{(1)}(m)}=JL(M-1)$ and
$\abs{\setW^{(2)}(m,l)}=J(L-1)$; replaced $I(Q,V')$ and
$I(Q_1,V'|Q_0)$ by their minima $I(Q,V)$ and respectively
$I(Q_1,V|Q_0)$ which correspond to the most slowly decaying terms; and
applied the Type Counting Lemma to $\abs{\rsfsV_n(Q)}$. Hence,
\iftwocolumn
\begin{multline*}
  \liminf_{n\to`8} -\frac1n \log `b(V,`Q,`F_b^c)\\
  \geq \abs{\min\Set*{I(Q,V)-R_M, I(Q_1,V|Q_0)}-R_J-R_L}^+\\
  = \abs{I(Q_1,V|Q_0) - R_J-R_L + \abs{I(Q_1V|Q_0) - R_M}^-}^+
\end{multline*}  
\else
\begin{align*}
  \liminf_{n\to`8} -\frac1n \log `b(V,`Q,`F_b^c)
  &\geq \abs{\min\Set*{I(Q,V)-R_M, I(Q_1,V|Q_0)}-R_J-R_L}^+\notag\\
  &= \abs{I(Q_1,V|Q_0) - R_J-R_L + \abs{I(Q_1V|Q_0) - R_M}^-}^+
\end{align*}  
\fi
because $\min\Set{a,b}= b + \min\Set{0,a-b}$.

\subsection{Exponent for Eve}
\label{sec:exponent-eve}

The exponent of $`b(V,`Q,`F_e^c)$ for Eve can be calculated
analogously. With MMI decoding $`F_e^c(m)\cap T_V(\RMC_{jlm})$ is the
set of $\Mz\in T_V(\RMC_{jlm})$ that has $I(\RMU_m \wedge \Mz)$ no
larger than $I(\RMU_{m'} \wedge \Mz)$ for some misleading codeword
$\RMU_{m'}$ where $m'\in M`/\Set{m}$. i.e.
\begin{multline*}
  `F_e^c(m)\cap T_V(\RMC_{jlm}) =\{\Mz\in T_V(\RMC_{jlm}) \cap
  T_{Q_1V'}(\RMU_{m'}): \iftwocolumn \\ else \fi
  m'\in M`/\Set{m},
  V'\in\setV_e(V)\}
\end{multline*}
where 
the set of problematic conditional types for Eve is
$\setV_e(V):=\Set{V'\in\setV(Q):I(Q_0,Q_1V')\geq I(Q_0,Q_1V)}$. By the
union bound,
\begin{align*}
  \abs*{`F_e^c(m)\cap T_V(\RMC_{jlm})}
  &\leq \smashoperator[l]{\sum_{V'\in\setV_e(V)}}
  \smashoperator[r]{\sum_{m'\in M`/\Set{m}}}
  \abs{T_V(\RMC_{jlm}) \cap
    T_{Q_1V'}(\RMU_{m'})}
\end{align*}
Since $\RMC_{jlm}$ is independent of $\RMU_{m'}$
where $m'\neq m$, the Packing Lemma~\ref{lem:packing} without
conditioning (but with $\hat{Q}$ assigned as $Q_0$, and $\hat{V}$
assigned as $Q_1V'$) gives, for all $n\geq n_0(`d,\abs\setU)$,
\iftwocolumn
\begin{align*}
  \opE`1(\tfrac{\abs{T_V(\RMC_{jlm}) \cap T_{Q_1V'}(\RMU_{m'})}}
  {\abs{T_V(\RMC_{jlm})}}`2)
  &\leq \exp\Set*{-n[I(Q_0,Q_1V')-`d]}
\end{align*}
\else
\begin{align*}
  \opE`1(\frac{\abs{T_V(\RMC_{jlm}) \cap T_{Q_1V'}(\RMU_{m'})}}
  {\abs{T_V(\RMC_{jlm})}}`2)
  &\leq \exp\Set*{-n[I(Q_0,Q_1V')-`d]}
\end{align*}
\fi
Substituting this into the previous inequality, we have for $n$
sufficiently large that,
\begin{align*}
  \opE`1(\frac{\abs*{`F_e^c(m)\cap T_V(\RMC_{jlm})}}{\abs*{ T_V(\RMC_{jlm})}}`2)
  &\leq M\exp\Set{-n[I(Q_0,Q_1V)-`d]}
\end{align*}
where we have replaced $I(Q_0,Q_1V')$ by it minimum $I(Q_0,Q_1V)$. The
exponent is therefore,
\begin{align*}
  \liminf_{n\to`8} -\frac1n \log `b(V,`Q,`F_e^c)
  &\geq \abs*{I(Q_0,Q_1V)-R_M}^+
\end{align*}
which completes the proof the Lemma~\ref{lem:wpack}

\section{Results}
\label{sec:result}

The exponents of $`b(V,`Q,`J)$, $`b(V,`Q,`F_b^c)$ and
$`b(V,`Q,`F_e^c)$ calculated in Lemma~\ref{lem:woverlap} and
Lemma~\ref{lem:wpack} using the random code in
Definition~\ref{def:randomseq} and the coding scheme in
Definition~\ref{def:achieving} give an initial set of lower bounds to the
exponents by Lemma~\ref{lem:E}. As discussed in
Section~\ref{sec:transm-junk-data}, the bounds can then be extended
with prefixed DMC $\tilde{V}$ by rewriting $(W_b,W_e)$ as $(\tilde V
W_b,\tilde V W_e)$.

To obtain the final version of the bounds, consider the following rate
reallocation: move the first $R\in[0,R_L]$ bits of the
secret to the end of the public message, and encode them with a
wiretap channel code at rate $(R_M+R,R_L-R)$.

\begin{thm}[Inner bound of achievable exponent triples]
  \label{thm:exponents}
  For every rate triple $(R_M,R_L,R_{`l})$, we have for all $R\in
  [0,R_L]$, $R_J\geq 0$, finite
  sets $\setU$ and $\tilde{\setX}$, distribution $Q:=Q_0\circ
  Q_1\;(Q_0\in\rsfsP(\setU),Q_1\in\rsfsP(\setX)^{\setU})$,
  transitional probability matrix $\tilde V \in
  \rsfsP(\setX)^{\setU\times \tilde \setX}$, the exponent triple
  $(E_b,E_e,S_e)$ satisfying the following is achievable (see
  Definition~\ref{def:achievable}) for the wiretap channel
  $\Set{W_b,W_e}$.
  \begin{align*}
    \begin{split}
      E_b &\geq \min_{V\in\rsfsP(\setY)^{\setU\times\setX}}
      D(V\|\tilde{V} W_b|Q) \\ &\quad
      + \lvert I(Q_1,V|Q_0) - R_J - R_L + R \\
      &\qquad +
        \abs*{I(Q_0,Q_1V) - R_M - R}^-\rvert^+
    \end{split}\\
    \begin{split}
      E_e &\geq \min_{V\in\rsfsP(\setZ)^{\setU\times\setX}}
      D(V\|\tilde{V} W_e|Q) \\ &\quad
      +\abs*{I(Q_0,Q_1 V) - R_M - R}^+
    \end{split}\\
    \begin{split}
      S_e &\geq \min_{V\in\rsfsP(\setZ)^{\setU\times\setX}}
      D(V\|\tilde{V} W_e|Q) \\ &\quad
      + \abs*{ R_L - R - R_{`l} +
        \abs*{R_J - I(Q_1,V|Q_0)}^-}^+
    \end{split}
  \end{align*}
\end{thm}

From this, we can compute an inner bound to the region of strongly
achievable rate triple for which above inner bound to the achievable
exponent triple are all strictly positive. To simplify notation, let
$(\RU,\tilde\RX,\RX,\RY,\RZ)$ be some random variables distributed as
$Q_0(u)Q_1(\tilde{x}|u){\tilde V}(x|u,\tilde{x})W_b(y|x)W_e(z|x)$.
(Note that $(\RU,\tilde{\RX})\to\RX\to\RY\RZ$.) Since information
divergence $D(V\|W)$ is zero at $V=W$ and positive otherwise, the
exponents are positive iff, for $R\in[0,R_L]$ and $R_J\geq 0$
\begin{subequations}
\label{eq:RC1}
\begin{align}
  R_J + R_L - R &< I(\tilde{\RX} \wedge \RY|\RU) \label{eq:R1}\\
  R_J + R_L + R_M & < I(\RU \tilde{\RX} \wedge \RY)\label{eq:R2}\\
  R_M + R & < I(\RU \wedge \RZ)\label{eq:R3}\\
  R_L - R & > R_{`l} \label{eq:R4}\\
  R_L -R + R_J &> R_{`l} + I(\tilde{\RX}\wedge \RZ|\RU)\label{eq:R5}
\end{align}
\end{subequations}
$R$ and $R_J$ can be eliminated without loss of optimality by the
Fourier-Motzkin elimination\cite{matousek2007} (see
Lemma~\ref{lem:FM}), which gives the following.

\begin{thm}[Inner bound of strongly achievable rate triples]
  \label{thm:rates}
  $(R_M,R_L,R_{`l})$ is strongly achievable for the wiretap
  channel $\Set{W_b:\setX\mapsto \setY,W_e:\setX\mapsto\setZ}$ if 
  \begin{subequations}
    \label{eq:rc}
    \begin{align}
      0\leq R_{`l} &< R_L \label{eq:r1}\\
      R_{`l} &< I(\tilde{\RX} \wedge \RY|\RU)
      - I(\tilde{\RX} \wedge \RZ|\RU)\label{eq:r2}\\
      0 \leq R_M &< I(\RU\wedge \RZ)\label{eq:r3}\\
      R_M + R_{`l}&<I(\RU\wedge\RY)+I(\tRX \wedge\RY|\RU)-I(\tRX\wedge \RZ|\RU)\label{eq:r4}\\
      R_M + R_L &< I(\tilde{\RX}\wedge \RY|\RU)
      +\min\Set{I(\RU\wedge \RY),I(\RU\wedge \RZ)}\label{eq:r5}\hspace*{-1em}
    \end{align}
  \end{subequations}
  for some $(\RU ,\tRX)\to \RX\to \RY\RZ$ with
  $P_{\RY|\RX}=W_b$ and $P_{\RZ|\RX}=W_e$. It is admissible to have
  $\RU$ as a deterministic function of $\tRX$ and 
  \begin{align*}
    \abs\setU &\leq 4+ \min\Set{\abs\setX -1,\abs\setY + \abs\setZ -2}\\
    \abs\tsetX 
    &\leq \abs\setU `1(2+ \min\Set{\abs\setX-1,\abs\setY+ \abs\setZ -2}`2)
  \end{align*}
  which implies $\RU\to \tRX\to\RX\to\RY\RZ$ and $I(\RU\tRX\wedge
  \RY)=I(\tRX\wedge\RY)$.
\end{thm}

The admissible constraints are obtained from \cite{csiszar1978} as
described in Lemma~\ref{lem:ac}. They can be imposed without changing
the inner bound. Example~\ref{eg:rates} illustrates how to compute an
inner bound of the achievable rate tuples using the Multi-Parametric
Toolbox\cite{mpt} in Matlab.

The closure of the rate region of $(R_M,R_L,R_{`l})$ is indeed
equivalent to the closure of the rate region of $(R_0,R_1,R_e)$ in
Theorem~1 of \cite{csiszar1978}. More precisely, we have the following.

\begin{pro}[Equivalent rate region]
  \label{pro:rates}
  Let $\setR$ be the inner bound of strongly achievable rate tuples
  $(R_M,R_L,R_{`l})$ in Theorem~\ref{thm:rates}, and $\setR'$ be the
  set of rate tuples that satisfies,
    \begin{subequations}
    \label{eq:rc'}
    \begin{align}
      0\leq R_{`l} &< R_L \label{eq:r'1}\\
      R_{`l} &< I(\tilde{\RX} \wedge \RY|\RU)
      - I(\tilde{\RX} \wedge \RZ|\RU)\label{eq:r'2}\\
      0 \leq R_M &< \min\Set{I(\RU\wedge \RY),I(\RU\wedge\RZ)}\label{eq:r'3}\\
      R_M + R_L &< I(\tilde{\RX}\wedge \RY|\RU)
      +\min\Set{I(\RU\wedge \RY),I(\RU\wedge \RZ)}\label{eq:r'5}\hspace*{-1em}
    \end{align}
  \end{subequations}
  for some $(\RU,\tRX)\to\RX\to\RY\RZ$ with the same admissible
  constraints as $\setR$. Then, $\setR=\setR'$. 
\end{pro}

Hence, $\setR$ is convex by Lemma~5 of \cite{csiszar1978} and the
closure of its projection on $(R_M,R_L)$ is the rate region for the
asymmetric broadcast channel by Corollary~5 of \cite{csiszar1978}.
Suppose $W_b$ is \emph{more capable}\cite{korner75} than $W_e$, i.e.
$I(\RX\wedge\RY)\geq I(\RX\wedge\RZ)$ for all
$P_{\RX}\in\rsfsP(\setX)$. Then it is admissible to have $\tRX=\RX$
(i.e. no prefix DMC) by a straightforward extension of the proof of
Theorem~3 in \cite{csiszar1978}. It also follows that $0\leq R_{`l} <
\max_{P_{\RX}}[I(\RX\wedge\RY)-I(\RX\wedge\RZ)]$ is the projection of
$\setR$ on $R_{`l}$. Assume the stronger condition that $W_b$ is
\emph{less noisy}\cite{korner75} than $W_e$, i.e. $I(\RU\wedge
\RY)\geq I(\RU\wedge \RZ)$ for any $\RU\to\RX\to\RY\RZ$. Then, by
Theorem~3 in \cite{csiszar1978}, it is addmissible to have $\RU$
deterministic in addition to $\tRX=\RX$ to obtain the projection on
$(R_L,R_{`l})$.

\begin{proof}[Proof of Proposition~\ref{pro:rates}]
  Without loss of generality, consider some $\RU\to\tRX\to\RX\to\RY\RZ$ with $\setU\cap \tsetX =
  `0$. Let $\RU_{`a}$ be a random variable such that it is $\tRX$ with
  probability $`a$ and $\RU$ with probability $1-`a$, and that
  $\ds1\Set{\RU_{`a}=\tRX}$ is independent of
  $(\RU,\tRX,\RX,\RY,\RZ)$.\footnote{This proof technique is from
    the proof of Theorem~4.1 in \cite[p.360]{csiszar1981}.} Then,
  $\RU_1=\tRX$, $\RU_0=\RU$, $\RU_{`a}\to\tRX\to\RX\to\RY\RZ$,
  \begin{align*}
    I(\RU_{`a} \wedge \RY)&=(1-`a)I(\RU\wedge
    \RY)+`aI(\tRX\wedge\RY)\\
    I(\tRX \wedge \RY|\RU_{`a})&= I(\tRX \wedge \RY)-I(\RU_{`a} \wedge \RY)
  \end{align*}
  and similarly for $\RZ$. Thus, we can define $\setR_{`a}$ and
  $\setR'_{`a}$ as the corresponding rate polytopes defined by the
  linear constraints in \eqref{eq:rc} and \eqref{eq:rc'} respectively.

  If we impose \eqref{eq:r'3} on $\setR_0$, the resulting polytope is
  the same as $\setR'_0$ because \eqref{eq:r3} and \eqref{eq:r4} are
  redundant under \eqref{eq:r2} and \eqref{eq:r'3}. Thus,
  $\setR_0\supset\setR'_0$, which implies $\setR\supset \setR'$.
  
  If $I(\RU\wedge\RZ)\leq I(\RU\wedge\RY)$, then \eqref{eq:r'3} is
  equivalent to \eqref{eq:r3}. By the previous argument,
  $\setR_0=\setR'_0$.

  If $I(\tRX \wedge \RY|\RU)\leq I(\tRX \wedge \RZ|\RU)$, then both
  $\setR_0=\setR'_0=`0$ by identical constraints \eqref{eq:r2} and
  \eqref{eq:r'2}.

  Consider $I(\RU \wedge \RY)\leq I(\RU \wedge\RZ) \leq
  I(\tRX\wedge\RZ)\leq I(\tRX\wedge\RY)$. Choose $`a$ such that
  $I(\RU_{`a} \wedge \RY)=I(\RU \wedge \RZ)$. The convex hull,
  $\op{Hull}(\setR'_0,\setR'_{`a})$, contains $\setR_0$ primarily
  because the hyperplane of \eqref{eq:r3} and \eqref{eq:r4} for
  $\setR_0$ intersects at,
  \begin{align*}
    R_{`l}&=I(\tRX\wedge\RY)-I(\tRX\wedge \RZ)\\
    &\leq I(\tRX\wedge\RY|\RU_{`a})-I(\tRX\wedge \RZ|\RU_{`a})
  \end{align*}
  which is contained by the half-space \eqref{eq:r'2} (with non-strict
  inequality instead) for $\setR'_{`a}$. This is illustrated in
  \figref{fig:rc1}.  For comparison, $\setR_0$ is plotted with blue
  dotted frame in each sub-figure. It is contained by the convex hull in
  \figref{fig:rc1c} as expected.

  \begin{figure}
    \centering
    \small

  \psfrag{s02}[rt][rt]{\color[rgb]{0,0,0}\setlength{\tabcolsep}{0pt}\begin{tabular}{r}$R_M$\end{tabular}}%
\psfrag{s03}[lt][lt]{\color[rgb]{0,0,0}\setlength{\tabcolsep}{0pt}\begin{tabular}{l}$R_L$\end{tabular}}%
\psfrag{s04}[b][b]{\color[rgb]{0,0,0}\setlength{\tabcolsep}{0pt}\begin{tabular}{c}$R_{\lambda}$\end{tabular}}%

\iftwocolumn 
\subfigure[$\setR'_0$]{
  \includegraphics[width=.45\mywidth]{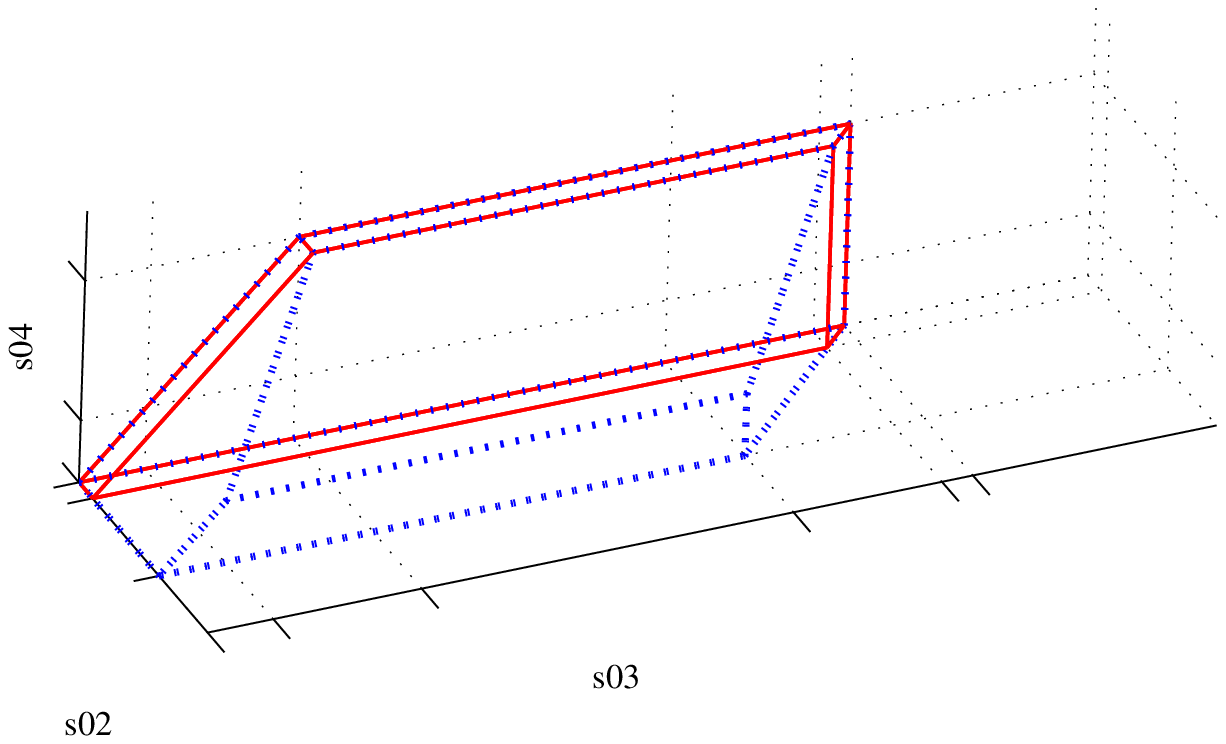}
}
\hfil
\subfigure[$\setR'_{`a}$]{
  \includegraphics[width=.45\mywidth]{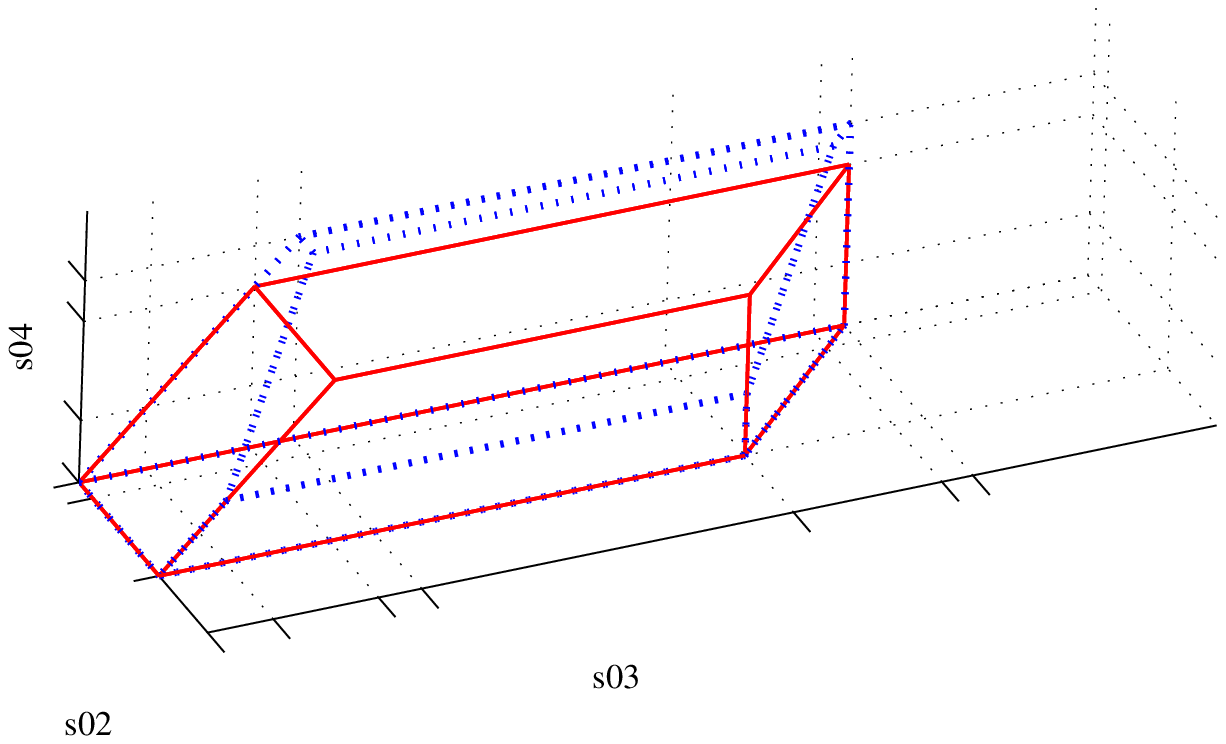}
}\\
\subfigure[$\op{Hull}(\setR'_0,\setR'_{`a})$]{
  \label{fig:rc1c}
  \includegraphics[width=.9\mywidth]{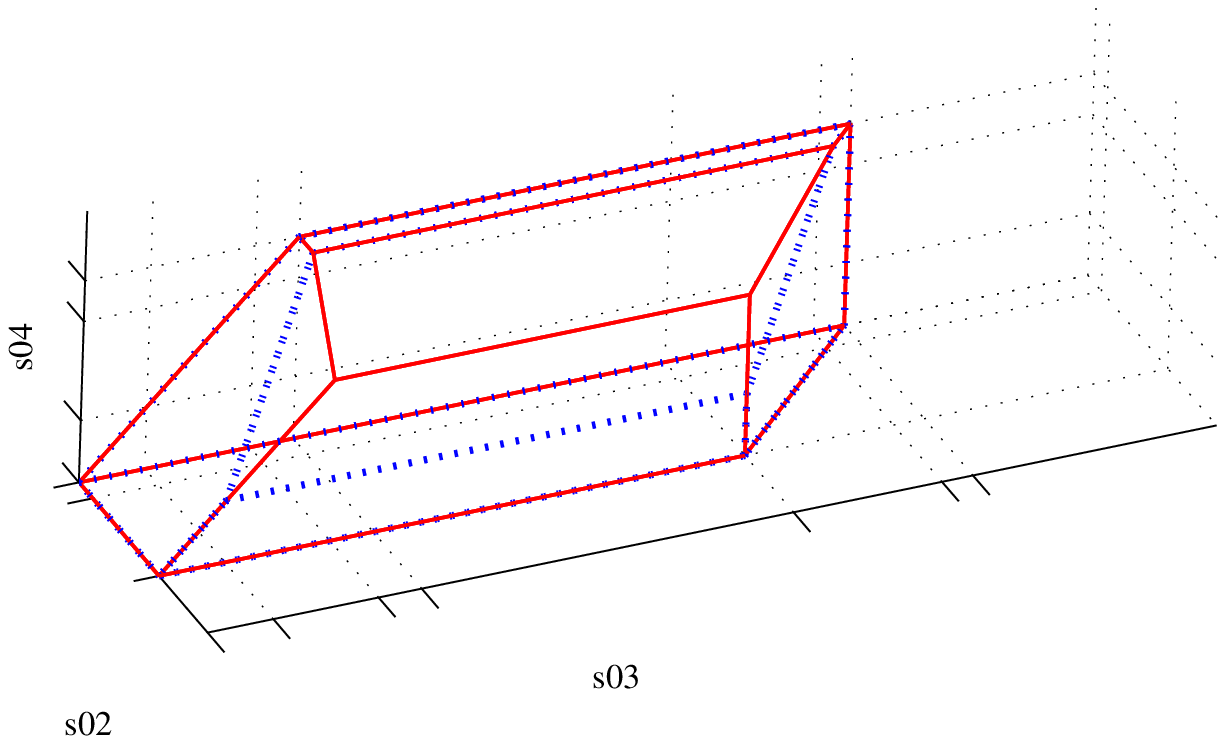}
}
\else
\subfigure[$\setR'_0$]{
  \includegraphics[width=.3\mywidth]{rc1a.eps}
}
\subfigure[$\setR'_{`a}$]{
  \includegraphics[width=.3\mywidth]{rc1b.eps}
}
\subfigure[$\op{Hull}(\setR'_0,\setR'_{`a})$]{
  \label{fig:rc1c}
  \includegraphics[width=.3\mywidth]{rc1.eps}
}
\fi
    \caption{$\setR_0\subset \op{Hull}(\setR'_0,\setR'_{`a})$ for the
      case $I(\RU \wedge \RY)\leq I(\RU \wedge\RZ) \leq
      I(\tRX\wedge\RY)$}
    \label{fig:rc1}
  \end{figure}
  \begin{figure}
    \centering
    \small

  \psfrag{s02}[rt][rt]{\color[rgb]{0,0,0}\setlength{\tabcolsep}{0pt}\begin{tabular}{r}$R_M$\end{tabular}}%
\psfrag{s03}[lt][lt]{\color[rgb]{0,0,0}\setlength{\tabcolsep}{0pt}\begin{tabular}{l}$R_L$\end{tabular}}%
\psfrag{s04}[b][b]{\color[rgb]{0,0,0}\setlength{\tabcolsep}{0pt}\begin{tabular}{c}$R_{\lambda}$\end{tabular}}%

\iftwocolumn 
\subfigure[$\setR'_0$]{
  \includegraphics[width=.45\mywidth]{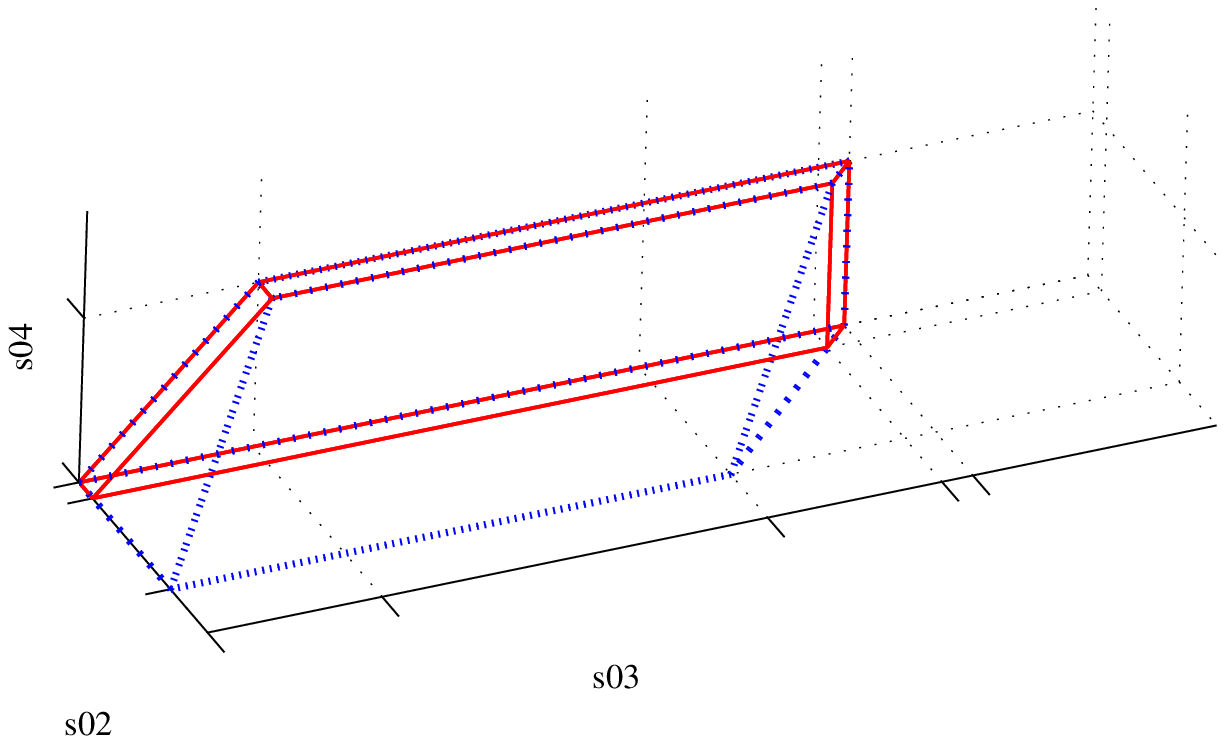}
}
\hfil
\subfigure[$\setR'_{`a}$]{
  \includegraphics[width=.45\mywidth]{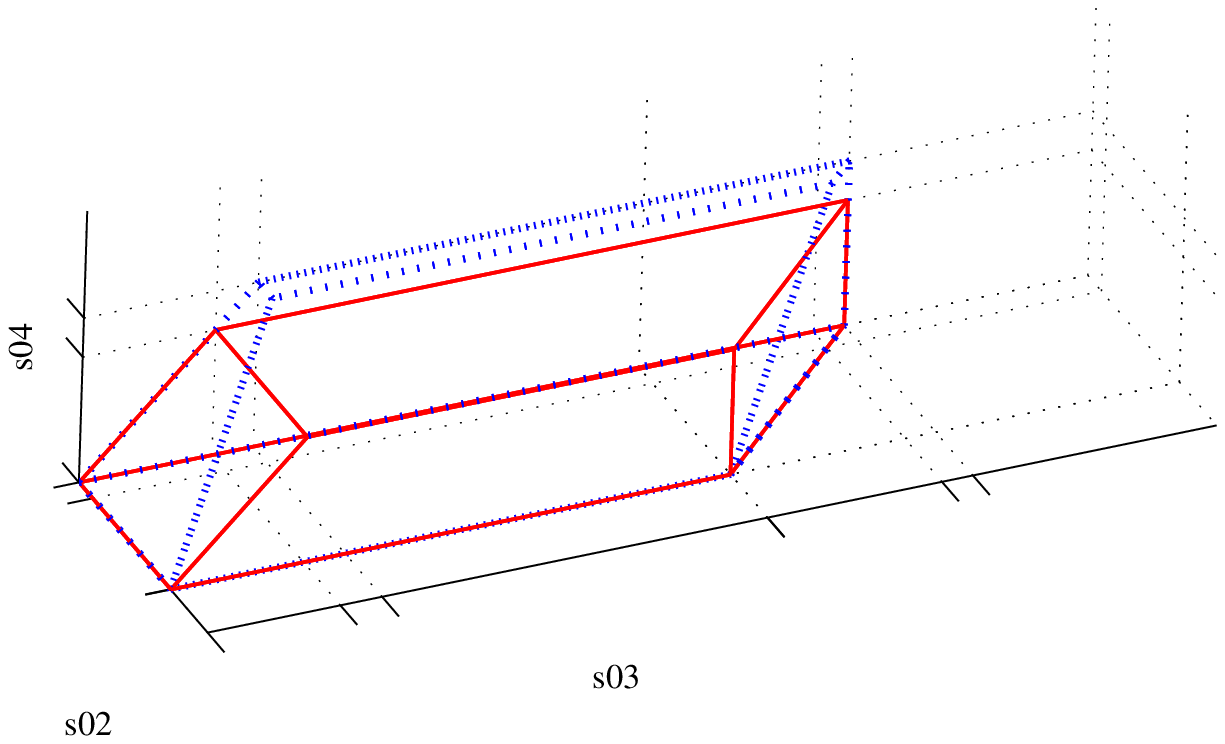}
}\\
\subfigure[$\op{Hull}(\setR'_0,\setR'_{`a})$]{
  \label{fig:rc2c}
  \includegraphics[width=.9\mywidth]{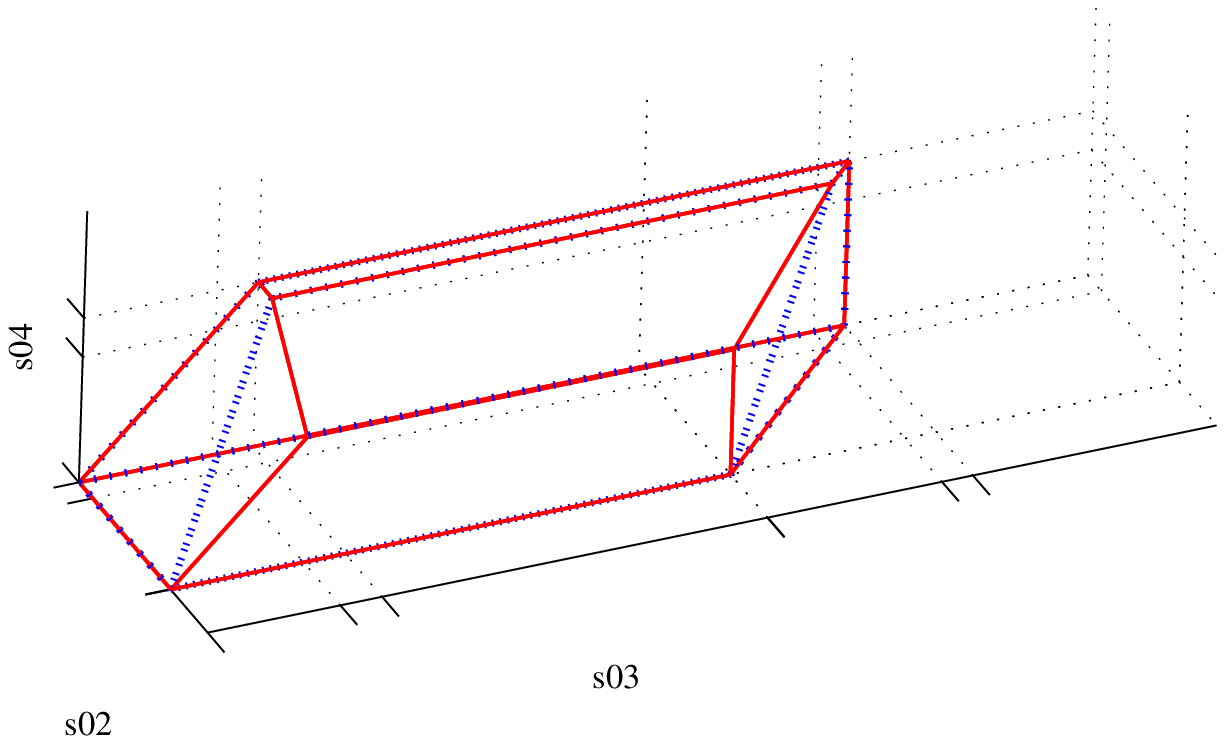}
}
\else
\subfigure[$\setR'_0$]{
  \includegraphics[width=.3\mywidth]{rc2a.eps}
}
\subfigure[$\setR'_{`a}$]{
  \includegraphics[width=.3\mywidth]{rc2b.eps}
}
\subfigure[$\op{Hull}(\setR'_0,\setR'_{`a})$]{
  \label{fig:rc2c}
  \includegraphics[width=.3\mywidth]{rc2.eps}
}
\fi
    
    \caption{$\setR_0\subset \op{Hull}(\setR'_0,\setR'_{`a})$ for the
      case $I(\tRX\wedge\RY) < I(\tRX \wedge\RZ)$}
    \label{fig:rc2}
  \end{figure}

  Finally, consider the case $I(\tRX\wedge\RY) < I(\tRX \wedge\RZ)$.
  Choose $`a$ such that $I(\RU_{`a}\wedge\RY)=I(\tRX\wedge\RY)-I(\tRX
  \wedge \RZ|\RU)$.\footnote{If $I(\tRX \wedge \RZ|\RU)=0$, choose
    $`a$ to approach $1$ from below to ensure that $\setR'_{`a}\neq
    `0$.} Then, $\op{Hull}(\setR'_0,\setR'_{`a})$
  contains $\setR_0$ primarily because the hyperplane of \eqref{eq:r4}
  intersects with the plane $R_{`l}=0$ at,
  \begin{align*}
    R_M &= I(\tRX\wedge\RY)-I(\tRX \wedge \RZ|\RU)
  \end{align*}
  which is contained by the half-space (with non-strict inequality) of
  \eqref{eq:r'3} for $\setR'_{`a}$. This is illustrated in
  \figref{fig:rc2}.
  Hence, we have $\setR_0$ a subset of
  $\op{Hull}(\setR'_0,\setR'_{`a})$ for some $`a\in(0,1)$, which
  implies $\setR\subset \setR'$ as desired.
\end{proof}

\section{Conclusion}
\label{sec:conclusion}

In doubt of a unifying measure of security, we have considered success
exponent as an alternative to equivocation rate for the wiretap
channel considered in \cite{csiszar1978}. We replace the maximal code
construction and typical set decoding in \cite{csiszar1978} with the
random coding scheme and maximum empirical mutual information decoding
in \cite{korner1980}. The lower bounds on the error exponents follow
from \cite{korner1980} with the well-known Packing Lemma (see
Lemma~\ref{lem:packing}), while the lower bound on the success
exponent is obtained with the approach of \cite{dueck1979} and a
technique we call the Overlap Lemma (see Lemma~\ref{lem:overlap}).
This lemma gives a doubly exponential behavior that enables us to
guarantee good realization of the random code for effective stochastic
encoding by transmission of junk data (see
Section~\ref{sec:transm-junk-data}). Combining with the prefix DMC
technique in \cite{csiszar1978} that adds artificial memoryless noise
to the channel input symbols, and a rate reallocation step of
transferring some secret bits to the public message before encoding
(see Section~\ref{sec:result}), we obtain the final inner bound of the
achievable exponent triples in Theorem~\ref{thm:exponents} with the
corresponding strongly achievable rate triples in
Theorem~\ref{thm:rates}. Proposition~\ref{pro:rates} shows that this
inner bound to the rate region is convex and coincides with the region
of achievable rate triples in Theorem~1 of \cite{csiszar1978}.

It is a straightforward extension to consider the maximum error
exponents and average success exponent over the messages. The same
bound follows by the usual expurgation argument and a more careful
application of the doubly exponential behavior of the Overlap Lemma.
Whether this tradeoff is optimal, however, is unclear. It would be
surprising if one can further improve the tradeoff by improving the
coding scheme.

\appendix
\section{Clarifications}
\label{sec:clarifications}

\begin{eg}[Maximum a priori and aposteriori success probability]
  \label{eg:perfect}
  Consider the following probability matrix,
  \begin{align*}
    P_{\RZ}&:= \bM \frac58 & \frac38 \eM\\
    P_{\RS|\RZ} &:= \bM  \frac45 & \frac15 \\ \frac23 & \frac13 \eM
  \end{align*}
  from which the a priori probability is $P_{\RS} = \bM \frac34 &
  \frac14 \eM$. Without knowing $\RZ$, Eve guesses $\RS$ successfully
  with probability at most $\frac34$ if one guess is allowed, and $1$
  if two guesses are allowed. If she knows $\RZ$, she still has the
  same maximum probability of success in each case because the most
  probable candidate for the secret is the same regardless of whether
  $\RZ$ is observed. Hence, Eve cannot achieve a better
  success probability regardless of $\RZ$, even though $\RZ$ is
  \emph{not} independent of $\RS$. Success probability fails to
  express the notion of perfect secrecy in this sense.
\end{eg}

\begin{eg}[Transmission of junk data]
  \label{eg:junkdata}
  Consider the case when there is no public message, and the coding is
  not restricted to constant composition code. \figref{fig:junkdata}
  illustrates the approach of transmission of junk data through a
  wiretap channel, which consists of a binary noiseless channel for
  Bob and a binary erasure channel for Eve.
  \begin{figure}[!h]
    \centering
    \input{fig_junkdata}
    \caption{An example of transmission of junk data}\label{fig:junkdata}
  \end{figure}
  While the channel input is perfectly observed by Bob, half of it is
  erased on average before it reaches Eve. Alice exploits this by
  sending one bit of junk $\RJ$ uniformly distributed in $\Set{0,1}$
  together with one bit of secret $l\in\Set{0,1}$ in two channel uses.
  The channel input is $\RMX=(\RJ,\RJ\oplus l)$ where $\oplus$ denotes
  the XOR operation.  Bob can recover the secret perfectly by the
  decoder $`f_b(\My):=y^{(1)}\oplus y^{(2)}$ since his observation
  $\RMY$ is equal to $\RMX$. Eve can use the same decoding if there is
  no erasure. However, if there is one or more erasures, her
  observation $\RMZ$ becomes independent of the secret, in which case
  she should uniformly randomly pick $0$ or $1$ as her guess to
  minimize the conditional error probability, provided that she can
  only make one guess.\footnote{We allow stochastic decoding here
    since the focus is the probability at block length $n=2$ instead
    of the exponent when $n\to`8$.} Thus, the conditional error
  probability is $0$ if there is no erasure, which happens with
  probability $1/4$, and $1/2$ otherwise. The overall conditional
  error probability is $3/8$.

  Note that if Alice uses a prefix DMC as described in
  Section~\ref{sec:transm-junk-data},
  Bob cannot achieve zero error
  probability. In other words, prefix DMC is strictly inferior in this
  case.\footnote{It would be more interesting to find an example in
    which prefix DMC is inferior even if Bob's probability of error
    cannot be made to $0$ by adding noise with memory like what the
    transmission of junk data does.}
\end{eg}

\begin{eg}[Prefix discrete memoryless channel]
  \label{eg:prefixDMC}
  Consider prefixing the wiretap channel
  $\Set*{W_b:\setX\mapsto\setY,W_e:\setX\mapsto\setZ}$ with the
  discrete memoryless channel $\Set{\tilde V}$ defined in
  \figref{fig:prefixDMC}. Each arrow connects an input alphabet to an
  output alphabet if the corresponding transition probability, labeled
  in the arrow, is non-zero.
  \begin{figure}[!h]
    \centering
    \input{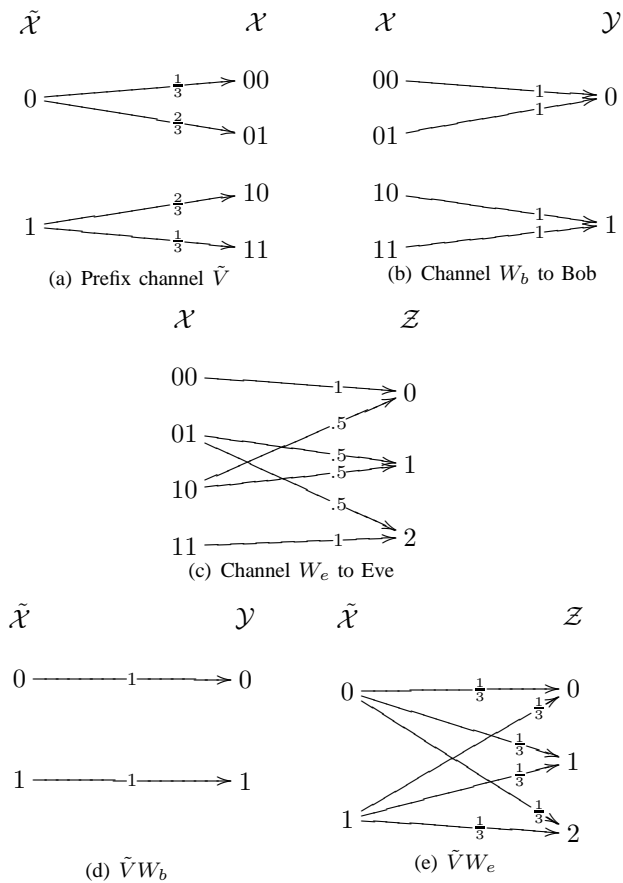}
    \caption{An example of prefix discrete
      memoryless channel}\label{fig:prefixDMC}
  \end{figure}
  Consider the case without prefixing the wiretap channel with $\tilde
  V$. Since $W_b$ is a weakly symmetric channel, the capacity is
  $1$~bit by the capacity formula for weakly symmetric
  in Theorem~8.2.1 of \cite{cover91eit}. Bob can achieve the capacity
  of $1$~bit with zero error probability and a single use of the
  channel iff Alice encodes $1$~bit of information using any of the
  following codebooks $`q(1):=\Set*{00,10}$, $`q(2):=\Set*{00,11}$,
  $`q(3):=\Set*{01,10}$ and $`q(4):=\Set*{01,11}$. If Alice wants to
  have zero error probability for Bob in $n$ channel uses with rate
  $n$~bits, the codebook has to be some concatenation of codebooks
  from $\Set{`q(i)}_{i=1}^4$. However, the channel input $\RX^n$ would
  not be independent of the channel output $\RZ^n$ to Eve. To argue
  this, consider the $i$-th channel use only. Suppose Alice uses
  $`q(1)$ to encode a uniformly random bit at that time slot. Then,
  given $\RZ^{(i)}=0$, we have $\RX^{(i)}=10$ with probability $2/3$
  rather than the prior probability $1/2$. The other cases can be
  argued similarly. In short, not randomizing over the code
  unavoidably leaks information to Eve. However, if the randomization
  is done by transmitting junk data, the useful data rate would drop
  below the capacity $1$~bit.
  
  Consider prefixing the wiretap channel with $\tilde V$. The prefixed
  channel $\tilde V W_b$ to Bob is a noiseless binary channel as shown
  in \figref{fig:prefixDMCd}. The prefixed channel $\tilde V W_e$
  to Eve, however, is completely noisy as shown in
  \figref{fig:prefixDMCe}. One can check that the channel output
  $\RZ$ is independent of $\RX$ for any input distribution on $\RX$.
  Thus, Alice can transmit at the capacity $1$~bit with zero error
  probability for Bob but without leaking any information to Eve.
  Prefixing discrete memoryless channel is strictly better than
  transmitting junk data in this case.
\end{eg}

\begin{lem}[random codeword]
  \label{lem:rcwd}
  For $`d>0$, $n\in `Z^+$, $Q:=Q_0\circ Q_1\;(Q_0\in \rsfsP_n(\setU),
  Q_1\in\rsfsV_n(Q_0,\setX))$, $V\in\rsfsV_n(Q,\setZ)$, $\Mu\in
  T_{Q_0}$, $n$-sequence $\RMX$ uniformly randomly chosen from
  $T_{Q_1}(\Mu)$, then
  \begin{align*}
    \Pr\Set{\Mz \in T_V(\Mu \circ \RMX)}
    &= \frac{\abs*{T_{\RX|\RU,\RZ}(\Mu,\Mz)}}{\abs*{T_{\RX|\RU}(\Mu)}}\\
    &\leq \exp\Set*{-n [I(Q_1,V|Q_0) - `d]}
  \end{align*}
  where the last inequality holds for all $n\geq
  n_0(`d,\abs\setU\abs\setX)$; $(\RU,\RX,\RZ)$ in the first equality
  is a random tuple with joint distribution $P_{\RU,\RX,\RZ}:=Q_0\circ
  Q_1\circ V$; $T_{P_{\RX|\RU,\RZ}}$ is denoted by $T_{\RX|\RU,\RZ}$
  and similarly for others; and $\abs*{T_{\RX|\RU,\RZ}(\Mu,\Mz)}$ with
  $(\Mu,\Mz)\in T_{\RU,\RZ}$ is denoted by $\abs*{T_{\RX|\RU,\RZ}}$.
\end{lem}

\begin{proof}
  Consider $\Mz \in T_{QV}$, for which the desired probability is non-zero.
  Since $\Mu\in T_{\RU}$, $\RMX\in T_{\RX|\RU}(\Mu)$, and $\Mz\in
  T_{\RZ}$, the event that $\Set{\Mz \in T_V(\Mu \circ \RMX)}$, or
  equivalently, $\Set{\Mz\in T_{\RZ|\RU,\RX}(\Mu\circ\RMX)}$, happens
  iff $(\Mu,\RMX,\Mz)\in T_{\RU,\RX,\RZ}$. This happens iff $\RMX\in
  T_{\RX|\RU,\RZ}(\Mu,\Mz)$. Hence, for all $\Mz\in T_{\RZ}$,
  \begin{align*}
    \Pr\Set*{\Mz \in T_V(\Mu \circ \RMX)}
    &= \Pr\Set*{\RMX\in T_{\RX|\RU,\RZ}(\Mu,\Mz)}\\
    &=
    \frac{\abs*{T_{\RX|\RU,\RZ}(\Mu,\Mz)}}{\abs*{T_{\RX|\RU}(\Mu)}}\\
    &\leq \exp\Set{-n[I(\RX \wedge \RZ|\RU)+`d]}
  \end{align*}
  where the last inequality is true for all $n\geq
  n_0(`d,\abs\setU\abs\setX)$ due to Lemma~1.2.5 of \cite{csiszar1981}
  that
  \begin{align*}
    \abs{T_{\RX|\RU,\RZ}(\Mu,\Mz)} &\leq \exp\Set{nH(\RX|\RU,\RZ)}\\
    \abs{T_{\RX|\RU}(\Mu)} &\geq (n+1)^{-\abs\setU\abs\setX}
    \exp\Set{nH(\RX|\RU)}
  \end{align*}
  Since $I(\RX \wedge \RZ|\RU)=I(Q_1,V|Q_0)$, this gives
  the desired bound.
\end{proof}

\begin{lem}
  \label{lem:choose}
  For all $n,\exp(nR),\exp(n`d)\in`Z^+$
  \begin{align*}
    {\exp(nR) \choose \exp(n`d)} \leq \exp\Set*{(\log e+n(R-`d))\exp(n`d)}
  \end{align*}
\end{lem}
\begin{proof}
  Let $a:=\exp(nR)$ and $b:=\exp(n`d)$. Then, we have the well-known
  inequality that ${a \choose b}\leq `1(\frac{a}{b}e`2)^b$, which
  gives the R.H.S.\ of the bound as desired. To derive this, note that
  $e^x \geq (1+x)$ for all $x\geq 0$. Thus,
  \begin{align*}
    e^{ax} & \geq (1+x)^a = \sum_{i=1}^a {a \choose i} x^i &\implies
    {a \choose b} &\leq e^{ax-b\ln x}
  \end{align*}
  Setting $x=b/a$ gives the desired inequality.
\end{proof}

\begin{lem}[list size constraint]
  \label{lem:list}
  For any subset $\setS\subset\setZ^n$ of observations and list decoder $`j$
  with list size $`l$, the corresponding decision region map
  $`J:L\mapsto 2^{\setZ^n}$ satisfies,
  \begin{align}
    \sum_{l\in L} \abs*{`J(l) \cap \setS} 
    &= `l \abs{\setS} \label{eq:list}
  \end{align}
\end{lem}

\begin{proof}
  The proof is by the \emph{double counting} principle,
    \begin{align*}
    \sum_{l\in L} \abs*{`J(l) \cap \setS} 
    &= \sum_{\Mz \in \setS } \sum_{l\in L} \ds1\Set{ l \in `j(\Mz)}
    = \sum_{\Mz \in \setS }`l
    = `l \abs\setS 
  \end{align*}
\end{proof}

\begin{lem}[Packing (with conditioning)]
\label{lem:packing}
Consider some finite sets $\setU$, $\setX$ and $\setY$, type
$Q_0\in \rsfsP_n(\setU)$, and canonical conditional types
$Q_1\in\rsfsV(Q_0,\setX)$ and $\hat{Q}_1\in\rsfsV(Q_0,\hat{\setX})$.
Let $Q:=Q_0\circ Q_1$ and $\hat{Q}:=Q_0\circ\hat{Q}_1$ be the
corresponding joint types; $\RMU$ be some random $n$-sequence
distributed over $T_{Q_0}$; $\RMX$ and $\hat{\RMX}$ be independently
and uniformly randomly drawn from $T_{Q_1}(\RMU)$ and
$T_{\hat{Q}_1}(\RMU)$ respectively; $\RMC:=\RMU\circ\RMX$ and
$\hat{\RMC}:=\RMU\circ\hat{\RMX}$ denote the element-wise
concatenations. Then, for all $`d>0$, $n\geq
n_0(`d,\abs\setU\abs\setX)$, $V\in \rsfsV_n(Q,\setY) \cap
\rsfsV_n(\hat{Q},\setY)$,
\begin{align*}
  \opE`1(
  \frac{\abs{T_V(\RMC) \cap T_{\hat V}(\hat \RMC)}}
  {\abs{T_V(\RMC)}}`2)
  &\leq \exp\Set*{-n[I(\hat{Q}_1,\hat V|Q_0)-`d]}
\end{align*}
\end{lem}

\begin{proof}
  Consider some realization $\Mu\in T_{Q_0}$ of $\RMU$. By conditional
  independence between $\RMX$ and $\hat{\RMX}$,
  \iftwocolumn
  \begin{multline*}
    \opE`1(\extendvert{
      \abs{T_V(\RMC) \cap T_{\hat V}(\hat \RMC)}
      |\RMU=\Mu}`2)\\
    \begin{aligned}
      &= \sum_{\My\in T_{Q_1 V}(\Mu)}
      \Pr\Set*{ \My \in T_V(\Mu \circ \RMX) }
      \Pr\Set*{ \My \in T_{\hat{V}}(\Mu \circ \hat{\RMX})
      }\\
      &\leq \sum_{\My\in T_{Q_1 V}(\Mu)} 
      \frac{\abs{T_{\RX|\RU,\RY}}}{\abs{T_{\RX|\RU}}}
      \exp\Set{-n[I(\hat{Q}_1,\hat{V}|Q_0) - `d]}\\
      &=
      \frac{\abs{T_{\RY|\RU}}\abs{T_{\RX|\RU,\RY}}}
      {\abs{T_{\RX|\RU}}}
      \exp\Set{-n[I(\hat{Q}_1,\hat{V}|Q_0) - `d]}
    \end{aligned}
  \end{multline*}
  \else
  \begin{align*}
    \opE`1(\extendvert{
      \abs{T_V(\RMC) \cap T_{\hat V}(\hat \RMC)}
      |\RMU=\Mu}`2)
    &= \sum_{\My\in T_{Q_1 V}(\Mu)}
    \Pr\Set*{ \My \in T_V(\Mu \circ \RMX) }
    \Pr\Set*{ \My \in T_{\hat{V}}(\Mu \circ \hat{\RMX})
    }\\
    &\leq \sum_{\My\in T_{Q_1 V}(\Mu)} 
    \frac{\abs{T_{\RX|\RU,\RY}}}{\abs{T_{\RX|\RU}}}
    \exp\Set{-n[I(\hat{Q}_1,\hat{V}|Q_0) - `d]}\\
    &=
    \frac{\abs{T_{\RY|\RU}}\abs{T_{\RX|\RU,\RY}}}
    {\abs{T_{\RX|\RU}}}
    \exp\Set{-n[I(\hat{Q}_1,\hat{V}|Q_0) - `d]}
  \end{align*}
  \fi
  where the first inequality follows from Lemma~\ref{lem:rcwd} (both
  the equality and inequality cases) $\forall n\geq
  n_0(`d,\abs\setU\abs\setX)$ with $\RU$, $\RX$ and $\RY$ and
  $T_{\RX|\RU,\RY}$ etc.\ defined analogously. Divide
  both sides by $\abs{T_V(\Mu\circ \RMX)}=\abs{T_{\RY|\RU,\RX}}$, and
  apply that fact that
  $\abs{T_{\RY|\RU}}\abs{T_{\RX|\RU,\RY}}=\abs{T_{\RX|\RU}}\abs{T_{\RY|\RU,\RX}}$,
  \iftwocolumn
  \begin{align*}
    \opE`1(\extendvert{
      \tfrac{\abs{T_V(\RMC) \cap T_{\hat V}(\hat \RMC)}}
      {\abs{T_{V}(\RMC)}}
      |\RMU=\Mu}`2)
    &\leq \exp\Set{-n[I(\hat{Q}_1,\hat{V}|Q_0) - `d]}
  \end{align*}
  \else
    \begin{align*}
    \opE`1(\extendvert{
      \frac{\abs{T_V(\RMC) \cap T_{\hat V}(\hat \RMC)}}
      {\abs{T_{V}(\RMC)}}
      |\RMU=\Mu}`2)
    &\leq \exp\Set{-n[I(\hat{Q}_1,\hat{V}|Q_0) - `d]}
  \end{align*}
  \fi
  Averaging both sides over $\RMU$ gives the desired bound.
\end{proof}

\begin{lem}[Fourier-Motzkin]
  \label{lem:FM}
  The rate constraints in \eqref{eq:RC1} with $R\in[0,R_L]$ and
  $R_J>0$ defines the same region of (non-negative) rate triples
  $(R_M,R_L,R_{`l})$ as the rate constraints in \eqref{eq:rc} do.
\end{lem}

\begin{proof}
  Consider applying the Fourier-Motzkin elimination. From
  \eqref{eq:RC1} and $R\in[0,R_L]$, we have,
  \begin{align*}
    -R &< 0\\
    -R + R_J + R_L &< I(\tilde{\RX} \wedge \RY|\RU)\\
    R - R_L & \leq 0\\
    R+ R_M & < I(\RU \wedge \RZ)\\
    R - R_L + R_{`l} &< 0\\
    R - R_J - R_L + R_{`l} &< - I(\tilde{\RX}\wedge \RZ|\RU)\\
    R_J + R_L + R_M & < I(\RU \tilde{\RX} \wedge \RY)
  \end{align*}
  Adding each of the first two inequalities to the next four eliminates $R$,
  which, together with $R_J\geq 0$,  gives,
  \begin{align*}
    -R_J & \leq 0\\
    - R_J - R_L + R_{`l} &< - I(\tilde{\RX}\wedge \RZ|\RU)\\
    R_J+R_L+ R_M & < I(\RU \wedge \RZ)+I(\tilde{\RX} \wedge \RY|\RU)\\
    R_J + R_{`l} &< I(\tilde{\RX} \wedge \RY|\RU)\\
    R_J + R_L + R_M & < I(\RU \tilde{\RX} \wedge \RY)\\
    R_M & < I(\RU \wedge \RZ)\\
    - R_L + R_{`l} &< 0\\
    R_{`l} &< I(\tilde{\RX} \wedge \RY|\RU)- I(\tilde{\RX}\wedge \RZ|\RU)
  \end{align*}
  where we have removed some inactive constraints.
  Adding each of the first two inequalities to the next three
  inequalities eliminates $R_J$, which gives \eqref{eq:rc} as desired.
\end{proof}

\begin{eg}[Inner bound of strongly achievable rate triples]
  \label{eg:rates}
  \lstset{language=Matlab,morekeywords={polytope,projection}}
  Consider the following wiretap channel and prefix DMC.
\begin{lstlisting}[linerange={1-6}]
% wiretap channel
p=0.1; PYX =[ 1-p p ; p 1-p; .5  .5]; % $P_{\RY|\RX}$
r=.4; PZX = [1 0; 1-r r; r 1-r]; % $P_{\RZ|\RX}$
% input distributions and prefix DMC
PX = [.25 .25 .5]; PtX = PX; PXtX= eye(3); % $P_{\RX}$, $P_{\tRX}$ and $P_{\RX|\tRX}$
q=.3; PUtX = [1-q q; 1 0 ; 0 1]; % $P_{\RU|\tRX}$
\end{lstlisting} 
  The prefix DMC is noiseless, i.e.\ $\RX=\tRX$. The channel and
  $\RU$ are constructed based on Counter-example~2 in
  \cite{korner75} with slight modifications.\footnote{This is such
    that the resulting constraints \eqref{eq:rc} on the rate region are
    not redundant for the purpose of illustration.} Define the Bayes' rule, conditional mutual information and
  entropy functions as follows.
\begin{lstlisting}
function PXY=bayes(PYX,PX) 
% compute $P_{\RX|\RY}$ from $P_{\RY|\RX}$ and $P_{\RX}$
PXY=repmat(PX,size(PYX,2),1).*PYX';
PXY=PXY./repmat(sum(PXY,2),1,size(PXY,2));
\end{lstlisting}
\begin{lstlisting}
function IQVP=I(Q,V,P) % compute $I(Q,V|P)$
if nargin<3
  P=ones(1,size(Q,1))./size(Q,1);
end
IQVP=H(Q*V,P)-H(V,P*Q);

function h=H(Q,P) % compute $H(Q|P)$
Q(Q==0)=1;
h=-P*sum(log2(Q).*Q,2);
\end{lstlisting}
  Then, the mutual information expressions required for the rate region can be
  computed as follows.
\begin{lstlisting}[linerange={8-13}]
% derived values
PU = PtX*PUtX; PtXU = bayes(PUtX,PtX); % $P_{\RU}$ and $P_{\tRX|\RU}$
PYtX=PXtX*PYX; PYU=PtXU*PYtX; % $P_{\RY|\tRX}$ and $P_{\RY|\RU}$
PZtX=PXtX*PZX; PZU=PtXU*PZtX; % $P_{\RZ|\tRX}$ and $P_{\RZ|\RU}$
IUY=I(PU,PYU); ItXYU=I(PtXU,PYtX,PU); % $I(\RU\wedge \RY)$ and $I(\tRX\wedge \RY|\RU)$
IUZ=I(PU,PZU); ItXZU=I(PtXU,PZtX,PU); % $I(\RU\wedge \RZ)$ and $I(\tRX\wedge \RZ|\RU)$
\end{lstlisting}
  Using the Multi-Parametric Toolbox\cite{mpt}, we first define the
  polytope satisfying the constraints from \eqref{eq:RC1} on
  $(R,R_J,R_M,R_L,R_{`l})$; and then project it to $(R_M,R_L,R_{`l})$,
  which should give the desired region in \eqref{eq:rc}.
\begin{lstlisting}[linerange={15-20}]
% constraints from \eqref{eq:RC1} on $(R,R_J,R_M,R_L,R_{`l})$
A=[-eye(5); -1 1 0 1 0; 1 0 0 -1 0; 1 0 1 0 0; 
  1 0 0 -1 1; 1 -1 0 -1 1; 0 1 1 1 0];
b=[zeros(1,5) ItXYU 0 IUZ 0 -ItXZU IUY+ItXYU]';
P=polytope(A,b);
R=projection(P,[3 4 5]); % Project to $(R_M,R_L,R_{`l})$ to obtain \eqref{eq:rc}
\end{lstlisting}
  Finally, plotting the region gives
  \figref{fig:region}.
\begin{lstlisting}[linerange={22-25}]
options.wire=1; plot(R,options); 
xlabel('R_M'); ylabel('R_L'); zlabel('R_{\lambda}');
set(gca,'CameraPosition',[1.5 -0.5 1],...
'CameraUpVector',[-0.5 0.2 0.8],'DataAspectRatio',[1 1 1]);
\end{lstlisting}
  \begin{figure}[!h]
    \centering
%
%
\begin{psfrags}%
\small
\psfragscanon%
%
\psfrag{s02}[rt][rt]{\color[rgb]{0,0,0}\setlength{\tabcolsep}{0pt}\begin{tabular}{r}$R_M$\end{tabular}}%
\psfrag{s03}[lt][lt]{\color[rgb]{0,0,0}\setlength{\tabcolsep}{0pt}\begin{tabular}{l}$R_L$\end{tabular}}%
\psfrag{s04}[b][b]{\color[rgb]{0,0,0}\setlength{\tabcolsep}{0pt}\begin{tabular}{c}$R_{\lambda}$\end{tabular}}%

\psfrag{rc1}[][]{\eqref{eq:r1}}
\psfrag{rc2}[][]{\eqref{eq:r2}}
\psfrag{rc3}[][]{\eqref{eq:r3}}
\psfrag{rc4}[][]{\eqref{eq:r4}}
\psfrag{rc5}[][]{\eqref{eq:r5}}

%
\psfrag{x01}[t][t]{0}%
\psfrag{x02}[t][t]{0.05}%
\psfrag{x03}[t][t]{0.1}%
%
\psfrag{v01}[r][r]{0}%
\psfrag{v02}[r][r]{0.05}%
\psfrag{v03}[r][r]{0.1}%
\psfrag{v04}[r][r]{0.15}%
\psfrag{v05}[r][r]{0.2}%
\psfrag{v06}[r][r]{0.25}%
\psfrag{v07}[r][r]{0.3}%
\psfrag{v08}[r][r]{0.35}%
%
\psfrag{z01}[r][r]{0}%
\psfrag{z02}[r][r]{0.05}%
\psfrag{z03}[r][r]{0.1}%
%
\includegraphics[width=.9\mywidth]{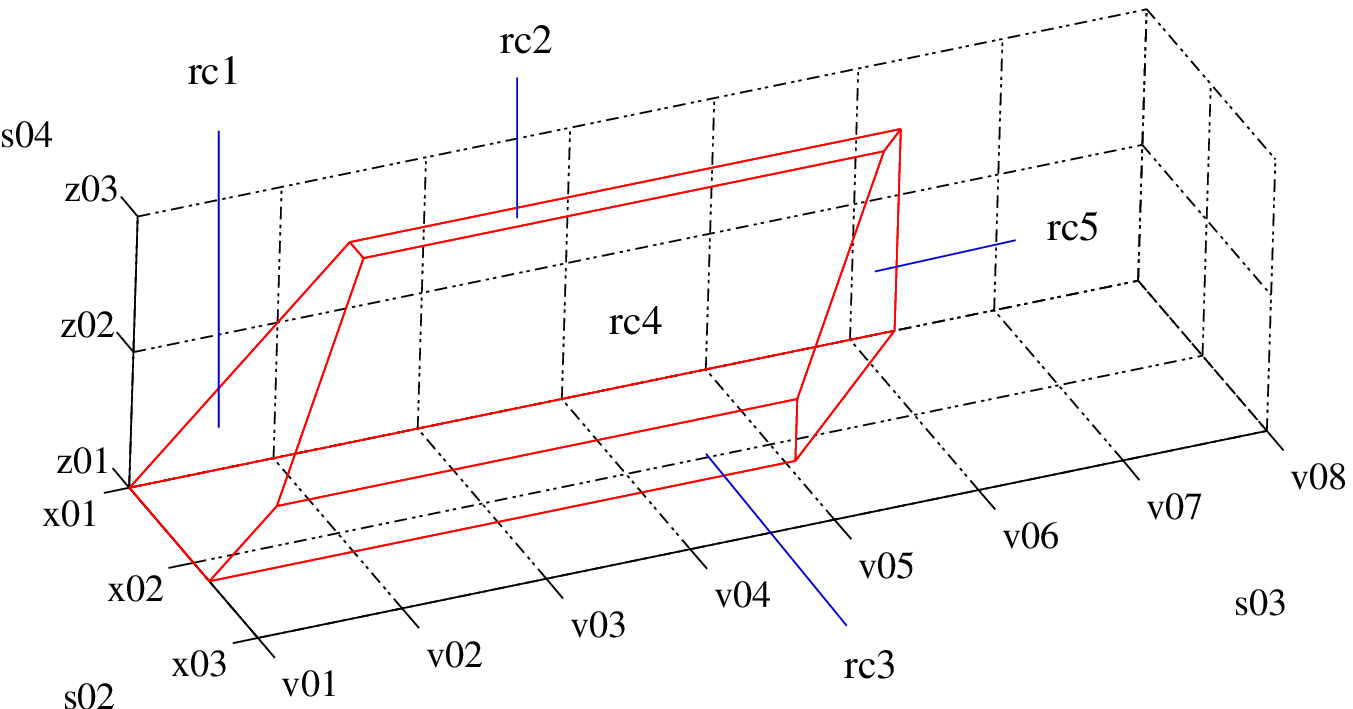}%
\end{psfrags}%
%

    \caption{An example of an inner bound to strongly achievable rate tuples}
    \label{fig:region}
  \end{figure}
  As expected, each facet corresponds to a constraint in
  \eqref{eq:rc}, indicated in the figure.
\end{eg}

\begin{lem}[admissible constraints]
  \label{lem:ac}
  Consider some random variables in the Markov chain $\RU'\tRX'
  \to \RX' \to \RY \RZ$ distributed over the finite sets $\setU'$,
  $\tsetX'$, $\setX$, $\setY$ and $\setZ$ respectively. Then there exists
  $\RU\to\tRX\to\RX \to \RY\RZ$ with,
  \begin{align*}
    P_{\RY|\RX}(y|x)&=P_{\RY|\RX'}(y|x) && ,\forall (x,y)\in
    \setX\times\setY\\
    P_{\RZ|\RX}(y|x)&=P_{\RZ|\RX'}(z|x) && ,\forall (x,z)\in
    \setX\times\setZ
  \end{align*}
  and
  \begin{subequations}
    \label{eq:I1}
    \begin{align}
      I(\RU \wedge \RY) &= I(\RU' \wedge \RY)\\
      I(\RU \wedge \RZ) &= I(\RU' \wedge \RZ)\\
      I(\tRX \wedge \RY|\RU) &= I(\tRX' \wedge \RY|\RU')\\
      I(\tRX \wedge \RY|\RU) &= I(\tRX' \wedge \RY|\RU')
    \end{align}    
  \end{subequations}
  and
  \begin{subequations}
    \label{eq:ac}
    \begin{align}
      \abs\setU &= 4+ \min\Set{\abs\setX -1,\abs\setY + \abs\setZ -2}\\
      \abs\tsetX 
      &= \abs\setU `1(2+ \min\Set{\abs\setX-1,\abs\setY+ \abs\setZ -2}`2)\\
      H(\RU|\tRX) &= 0
    \end{align}
  \end{subequations}
  Furthermore, $\RX=\RX'$ if $\abs{\setX}-1\leq \abs\setY+\abs\setZ-2$.
\end{lem}

\begin{proof}
  Since the following proof is a minor extension to
  \cite[(A.22)]{csiszar1978}, we will give only the changes
  as follows. Readers should refer to \cite{csiszar1978} for details.

  With $\tRX'':=(\RU,\tRX')$, we have $I(\tRX''\wedge
  \RY|\RU)=I(\tRX'\wedge \RY|\RU)$ and similarly for $I(\tRX''\wedge
  \RZ|\RU)$. It suffices to show the desired existence with
  $\tRX'$ replaced by $\tRX''$ on the R.H.S.\ of \eqref{eq:I1}.

  Consider the case $\abs{\setX}-1\leq \abs\setY+\abs\setZ-2$. The
  admissible constraint \eqref{eq:ac} is equivalent to
  \cite[(A.22)]{csiszar1978}. (n.b.\ $\setV$ in \cite{csiszar1978} is
  $\tsetX$ here.) 
  The proof therein also implies $\RX=\RX'$.
  because $(\RX',\RY,\RZ)$ need not be changed.

  Suppose $\abs{\setX}-1> \abs\setY+\abs\setZ-2$ instead. To achieve
  $H(Y)$ and $H(Z)$ in \cite[(A.24), (A.25)]{csiszar1978}, one
  can replace (A.23) by
  \begin{equation}
    \label{eq:I2}
  \begin{aligned}
    \Pr(Y=y) = \sum_{u\in\setU} \Pr\Set{U=u} f_y(\bar{\Mp}_u)\\
    \Pr(Z=z) = \sum_{u\in\setU} \Pr\Set{U=u} f_z(\bar{\Mp}_u)
  \end{aligned}
  \end{equation}
  where, using the notation in \cite{csiszar1978},
  \begin{align*}
    f_y(\bar{\Mp}):= \bar{\Mp}^Y(y)\quad\text{and}\quad
    f_z(\bar{\Mp}):= \bar{\Mp}^Z(z)
  \end{align*}
  Only $\abs{\setY}-1$ of the functions $f_y(\bar{\Mp})$ and
  $\abs\setZ-1$ of the functions $f_z(\bar{\Mp})$ are considered.
  Thus, as a consequence of the Eggleton-Carath\'eodory Theorem, $U$
  takes at most $(\abs\setY+\abs\setZ-2)+4$ different values to
  preserve (A.24) to (A.27) in \cite{csiszar1978} and \eqref{eq:I2}
  defined above.  Similarly, (A.28) can be replaced by the
  corresponding expressions on $\Pr(Y=y|U=u)$ and $\Pr(Z=z|U=u)$. For
  every fixed $u$, there exists a random variable $V_u$ with no more
  than $(\abs\setY+\abs\setZ-2)+2$ values preserving the set of
  desired equalities. With $\tRX$ here playing the role of the new $V$ in
  \cite{csiszar1978}, \eqref{eq:ac} follows.
 \end{proof}

\bibliography{main}
\bibliographystyle{plainurl}

\end{document}